\documentclass[iop,useAMS,usenatbib,preprint]{emulateapj}
\usepackage{amsmath}
\usepackage{graphicx}

\shorttitle{Two-stage Fragmentation}
\shortauthors{Bailey \& Basu}

\begin{document}

\title{Non-Ideal Magnetohydrodynamic Simulations of the Two-Stage Fragmentation Model for Cluster Formation}

\author{Nicole D. Bailey\altaffilmark{1,2} and Shantanu Basu\altaffilmark{1}}
\altaffiltext{1}{Department of Physics and Astronomy, University of Western Ontario, 1151 Richmond Street, London, Ontario, Canada, N6A 3K7}
\altaffiltext{2}{Current Address: School of Physics and Astronomy, University of Leeds, Leeds, United Kingdom, LS2 9JT}
\email{N.Bailey@leeds.ac.uk (NDB); basu@uwo.ca (SB)}

\begin{abstract}

We model molecular cloud fragmentation with thin disk non-ideal magnetohydrodynamic simulations that include ambipolar diffusion and partial ionization that transitions from primarily ultraviolet dominated to cosmic ray dominated regimes. These simulations are used to determine the conditions required for star clusters to form through a two-stage fragmentation scenario. Recent linear analyses have shown that the fragmentation length and time scales can undergo a dramatic drop across the column density boundary that separates the ultraviolet and cosmic ray dominated ionization regimes. As found in earlier studies, the absence of an ionization drop and regular perturbations leads to a single-stage fragmentation on parsec scales in transcritical clouds, so that the nonlinear evolution yields the same fragment sizes as predicted by linear theory. However, we find that a combination of initial transcritical mass-to-flux ratio, evolution through a column density regime in which the ionization drop takes place, and regular small perturbations to the mass-to-flux ratio are sufficient to cause a second stage of fragmentation during the nonlinear evolution. Cores of size $\sim 0.1$ pc are formed within an initial fragment of $\sim$ pc size. Regular perturbations to the mass-to-flux ratio also accelerate the onset of runaway collapse.

\end{abstract}


\section{Introduction}

Molecular clouds are the birthplaces of stars, and generally contain numerous star-forming condensations as well as rarefied regions devoid of star formation \citep[e.g.,][]{Onishi2002, Kirk2006, gol08}. Although characterized by a continuous range of sizes, the condensations can be loosely categorized as consisting of parsec-scale cluster-forming clumps and dense cores of $\sim 0.1$ pc scale that lead to single or weak multiple star systems. Dense cores show evidence of systematic inward gravitational collapse \citep{LM2011}, consistent with their association with single or small multiple systems. There is no evidence for continuous hierarchical fragmentation within cores, as originally envisioned by \citet{Hoyle1953}. These observations are consistent with the opposite idea of single-stage fragmentation \citep[e.g.,][]{Mouschovias1977}. However, the formation of the larger clumps that contain the cores may represent an earlier stage of fragmentation, leading to the idea of {\it two-stage fragmentation}. Recent observations reveal  subfragmentation within clumps. For example, early submillimeter observations of the B1 region of the Perseus molecular cloud revealed a clump to the east, B1-E \citep{Kirk2006, Jorgensen2007}, and more recent observations from the Herschel Gould Belt survey  and Green Bank Telescope (GBT) reveal the presence of previously unobserved core size fragments within B1-E \citep{Sadavoy2012}. 

In a previous paper, \citet[][hereafter BB12]{BB2012}, we presented a scenario for a two-stage fragmentation process in a molecular cloud. We showed through results of linear analysis of a partially ionized thin sheet that transcritical gas with visual extinction ($A_{V}$) between 1 and 4 magnitudes can form parsec scale clumps while the denser ($A_{V} > 5$ mag) transcritical to supercritical gas can form smaller subparsec size cores. The lower column density (and $A_{V}$) regions have partial ionization determined by background ultraviolet starlight and the higher density regions have a lower ionization fraction determined by cosmic rays. These arguments were based on snapshots of physical conditions in molecular clouds, and not on time dependent models. The overall idea is that molecular clouds are assembled from HI cloud material that is generally subcritical \citep{hei05}, and that a cloud may resist fragmentation (due to the very long  ambipolar diffusion time) until flows primarily along the magnetic field lines raise the mass-to-flux ratio to slightly above the critical value. At this point, the transcritical cloud may form large fragments as predicted by the linear theory. As these fragments develop, they will also become more supercritical, since their evolution is partially driven by neutral-ion drift.  Once the column density in the fragments also crosses the column density threshold for the transition to the lower level of cosmic ray dominated ionization,  the fragmentation length and time scales will drop dramatically, and a second stage  of fragmentation may be possible.

In this paper, we employ time-dependent non-ideal magnetohydrodynamic (MHD) simulations including both the regime of ultraviolet starlight dominated ionization and the subsequent cosmic ray dominated regime at higher column densities. With this kind of model we can test the relevance of the two stage-fragmentation scenario in a time-dependent and nonlinear environment. We also include the effect of ongoing small amplitude perturbations to the cloud in order to follow the formation of parsec size clumps and the possible subsequent formation of subparsec size fragments. Similar non-ideal MHD simulations have been performed by \citet{Basu2009b} using the same numerical code. Those simulations modeled only the cosmic-ray dominated regime of partial  ionization, hence should be applied to a single stage of fragmentation that takes place at $A_{V} \gtrsim$ 5 mag.

The thin sheet approximation is an idealization to highly structured regions of molecular clouds, but has the important property of preferred length scales
for gravitational fragmentation, unlike a uniform medium. A dense layer may also be considered the region where turbulence remains low, while large amplitude turbulent motions continue in a rarefied surrounding medium. Numerical models of turbulent wave propagation in a stratified medium \citep{kud03, kud06, Pinto2012} support such a view.

The remainder of the paper is organized as follows. Section 2 describes the physical model and assumptions. Section 3 describes the model parameters explored in this study while Section 4 describes the simulations and results. In Section 5, we analyze our simulations in the context of the two-stage fragmentation model. Finally, a discussion and summary is presented in Section 6.

\section{Numerical Model}

We explore clump/core formation within partially ionized, isothermal, magnetic interstellar molecular clouds. Our model assumes planar clouds with infinite extent in the $x$- and $y$-directions and a local vertical half thickness $Z$. A full description of the assumptions, nonaxisymmetric equations and formulations can be found in \citet{BC2004, CB2006, Basu2009a, Basu2009b}, however we highlight those essential for this analysis below.

Our model includes the effect of ambipolar diffusion, a measure of the coupling of neutral particles with the magnetic field via ions bound to the field. This coupling is quantified by the time scale for collisions between neutrals and ions,
\begin{equation} 
\tau_{ni} = 1.4 \left(\frac{m_i +m_{H_2}}{m_i} \right) \frac{1}{n_i\langle\sigma w\rangle_{iH_2}}, 
\end{equation} 
where $m_{i}$ is the ion mass, $m_{H_2}$ is the mass of molecular hydrogen, $n_{i}$ is the number density of ions, and $\langle\sigma w\rangle_{iH_2}$ is the neutral-ion collision rate. Typical ions within a molecular cloud include singly ionized Na, Mg, and HCO which have a mass of 25 amu. Assuming collisions between H$_{2}$ and HCO$^+$, the neutral-ion collision rate is $1.69\times 10^{-9}$ cm$^{-3}$ s$^{-1}$ \citep{MM1973}.

The threshold for collapse within a molecular cloud is regulated by the normalized mass-to-flux ratio of the background reference state,
\begin{equation}
\mu_{0} \equiv 2\pi G^{1/2}\frac{\sigma_{n,0}}{B_{\rm ref}},
\end{equation}
where $(2\pi G^{1/2})^{-1}$ is the critical mass-to-flux ratio for gravitational collapse in the adopted model \citep{CB2006}, $\sigma_{n,0}$ is the initial column density and $B_{\rm ref}$ is the magnetic field strength of the background reference state. Supercritical, transcritical and subcritical regions are those with mass-to-flux ratios greater than, approximately equal to, and less than one, respectively. In the limit where $\tau_{ni} \rightarrow 0$, the medium is defined to be flux-frozen, that is, frequent collisions between the neutral particles and ions couples the neutrals to the magnetic field. Under these conditions, subcritical regions are supported by the magnetic field and only supercritical regions may collapse within a finite time frame. Non-zero values of $\tau_{ni}$ are inversely dependent on the ion number density and therefore on the degree of ionization for a fixed neutral density.

Our model is characterized by several dimensionless free parameters including a dimensionless form of the initial neutral-ion collision time ($\tau_{ni,0}/t_{0}~\equiv~2\pi G\sigma_{n,0}\tau_{ni,0}/c_{s}$) and a dimensionless external pressure ($\tilde{P}_{\rm ext} \equiv 2 P_{\rm ext}/\pi G \sigma^{2}_{n,0}$).  Here, $c_{s} = (k_{B} T/m_{n})^{1/2}$ is the isothermal sound speed; $k_{B}$ is the Boltzmann constant, $T$ is the temperature in Kelvin, and $m_{n}$ is the mean mass of a neutral particle ($m_{n} = 2.33$ amu). We normalize column densities by $\sigma_{n,0}$, length scales by $L_{0} = c_{s}^{2}/2\pi G \sigma_{n,0}$ and time scales by $t_{0} = c_{s}/2\pi G \sigma_{n,0}$. Based on these parameters, typical values of the units used and other derived quantities are 

\begin{eqnarray}
\nonumber\sigma_{n,0} &=& \frac{3.63\times 10^{-3}}{(1+\tilde{P}_{\rm ext})^{1/2}}\left(\frac{n_{n,0}}{10^3 \rm ~cm^{-3}}\right)^{1/2}\left(\frac{T}{10 ~\rm K}\right)^{1/2} \rm g~cm^{-2},\\ 
&&\\
c_{s} &=& 0.188\left(\frac{T}{10 ~\rm K}\right)^{1/2} \rm km~s^{-1},\\
t_{0} &=& 3.98\times 10^5\left(\frac{10^3 \rm~ cm^{-3}}{n_{n,0}}\right)^{1/2}(1 + \tilde{P}_{\rm ext})^{1/2}~\rm yr,\label{time}\\ 
\nonumber L_{0} &=& 7.48\times 10^{-2} \left(\frac{T}{10 ~\rm K}\right)^{1/2}\times\\
&&\left(\frac{10^3 \rm ~cm^{-3}}{n_{n,0}}\right)^{1/2}(1 + \tilde{P}_{\rm ext})^{1/2}~\rm pc,\label{length}
\end{eqnarray}
where $n_{n,0}$ is the initial neutral number density. For our analysis, we assume a dimensionless external pressure $\tilde{P}_{\rm ext} = 0.1$ and a temperature $T = 10$ K.

To model the fragmentation of a molecular cloud and the subsequent evolution of the substructures formed, we utilize the IDL simulation code developed by \citet{BC2004} \citep[see also][]{CB2006, Basu2009b}. This multi-fluid non-ideal MHD code solves the nonaxisymmetric MHD equations numerically in $(x,y)$ coordinates. This code employes the numerical method of lines \citep{Schiesser1991}, that is, the first order partial differential equations are converted into a set of coupled ordinary differential equations (ODEs) in time, with one ODE for each physical variable at each cell \citep{Basu2009b}. Time evolution of the system of ODEs is performed by using an Adams-Bashforth-Moulton predictor-corrector subroutine \citep{Shampine1994}. The code assumes an isothermal gas for the cloud. For early stages of star formation where the densities are below $10^{10}$ cm$^{-3}$, this is a reasonable assumption \citep{Gaustad1963, Hayashi1966, Larson1969}. However, \citet{Zucconi2001} found that small departures from isothermality through a decrease in temperature could be expected in pre-protostellar cores for densities above 10$^{5}$ cm$^{-3}$. Above this threshold the gas temperature is expected to follow the dust temperature \citep{Zucconi2001}. Our simulations remain below the 10$^{5}$ cm$^{-3}$ threshold, allowing us to assume that the cloud evolves isothermally. 

This numerical code has been used previously to investigate the effect of ambipolar diffusion on the fragmentation length and time scales within a molecular cloud \citep[see][]{BC2004,CB2006,Basu2009b}. These previous investigations all assume ionization of the medium via cosmic rays and impose an ionization profile with an initial neutral-ion collision time of $\tau_{ni,0} = 0.2t_{0}$. Here, we investigate the two-stage fragmentation model which assumes a step-like ionization profile that includes both the ultraviolet (UV) and cosmic ray (CR) regimes. With the inclusion of this ionization profile, the ionization fraction regime (and neutral-ion collision time) within the cloud depend on the column density/visual extinction of each region and therefore changes as the cloud evolves. To allow for this variation with respect to column density/visual extinction, the ionization fraction of each pixel is calculated based upon the column density at each time step. We assume the same ionization profile as BB12\footnote{In BB12, the exponent $-1/2$ for the $(1-\tilde{P}_{\rm ext})$ term is missing a negative sign in Equation (24) and $\log \chi_{i,c}$ should be -7.362 to achieve a smooth transition. These small errors do not affect the two-stage fragmentation model presented in BB12.}, i.e.,
\begin{equation} 
\log \chi_{i} = \left\{ \begin{array}{ll}
\log \chi_{i,0} + 0.5(\log \chi_{i,c} - \log \chi_{i,0})\times& \\
~~~~~~\left(1 + \tanh\frac{A_{V}-A_{V, \rm crit}}{A_{V,d}}\right) & \mbox{$A_{V}\le A_{V,CR}$} \\
\log[ 1.148\times10^{-7}(1 +\tilde{P}_{\rm ext})^{-1/2}\times  &\\
~~~~~~\left(\frac{T}{10~\rm K}\right)^{1/2}\left(\frac{2.75~ \rm mag}{A_{V}}\right)] & \mbox{$A_{V} > A_{V,CR}$ },
\end{array}
\right.
\label{eqn:ionmodel}
\end{equation}
where $\chi_{i}$ is the ionization fraction, and $A_{V,CR} = 6.365$ mag is the location of the transition from the UV regime to the CR regime. The step function parameters are set to $\log \chi_{i,0} = -4.0$, $\log \chi_{i,c} = -7.362$, $A_{V,\rm crit} = 4.0$ mag, and $A_{V,d} = 1.05$ mag. Three decimal place accuracy is used for these transition parameters to ensure matching values of $\chi_{i}$ as well as a visually smooth transition of its slope, rather than for empirical reasons. The derived neutral-ion collision time for each pixel then depends on which ionization regime it is in. For the UV regime, the neutral-ion collision time is computed via 
\begin{eqnarray}
\nonumber\tau_{ni}/t_{0} = &0.262& \left(\frac{T}{10~\rm K}\right)^{1/2}\left(\frac{0.01~\rm g~cm^{-2}}{\sigma_{n,0}}\right)\times \\
&& \left(\frac{10^{-7}}{\chi_{i}}\right)(1+\tilde{P}_{\rm ext})^{-1}.
\label{eqn:tauni}
\end{eqnarray}
For the CR regime, we set the dimensionless neutral-ion collision time to $\tau_{ni,0}/t_{0} = 0.2$.

\section{Model Parameters}

We ran several simulations to test various realizations within the parameter space. Unlike previous studies using this code \citep{CB2006,Basu2009b}, the addition of the step-like ionization profile requires that the background column density within the simulations be given a dimensional value rather than be simply normalized to be unity. Since the aim of this investigation is to test the two-stage fragmentation model, we require an initially diffuse cloud and as such set the background column density to correspond to a visual extinction $A_{V,0} = 1$ mag. Using the prescription of \citet{Pineda2010} \citep[see also][]{BB2012} and assuming a mean molecular weight of 2.33 amu, the resulting conversion between visual extinction and mass column density is 
\begin{equation} 
\sigma_{n} =3.638\times 10^{-3} (A_{V}/\rm mag)~\rm g~cm^{-2}.
\label{av2sigma}
\end{equation}

All simulations begin with an initial linear column density perturbation, $\delta\sigma_{n}/\sigma_{n,0}$ which is a normally distributed random variable with mean equal to zero and standard deviation $A$. From here on, the value $A$ will be referred to as the amplitude. The random value of $\delta\sigma_{n}/\sigma_{n,0}$ for each pixel is then added to the background column density in that pixel. For some simulations, subsequent perturbations are applied at specific intervals ($\Delta t_{sp}/t_{0}$). We add only small-amplitude column density perturbations, corresponding to subsonic turbulence that is observed within dense cores of molecular clouds \citep[e.g.,][]{BM1989}. We do not add perturbations corresponding to supersonic turbulence that is inferred on large scales of molecular clouds \citep[e.g.,][]{Solomon1987}.  Our idea here is to see if the two-stage fragmentation can emerge naturally in an environment with small amplitude perturbations. Models with highly nonlinear turbulent forcing are left for another study.

All simulations are performed on a 512 $\times$ 512 periodic box. The box size is 64$\pi L_{0}$. This translates to a size of 15.16 pc, or a pixel size of 0.0296 pc, for $T$ = 10 K and $\sigma_{n,0} = 3.638\times 10^{-3}$ g~cm$^{-2}$. Finally, in order to compare the differences in the outcomes between the different models, we have fixed the random number generator to give the same realizations of the perturbations for each event i.e., the initial perturbation in all simulations is generated by the same seed. Therefore, any differences detected between the models are due to actual changes in the parameters and not based upon random stochastic events. However, maintaining the same seed for ongoing perturbations in the same model can artificially favour structure in certain regions. To avoid this, all subsequent perturbations in a single model are produced using the next sequential seed realization. All simulations are set to stop when $\sigma_{n}/\sigma_{n,0} \ge 10$ within any pixel. The initial parameters for the specific models can be found in Table~\ref{models}. The values quoted for the initial neutral-ion collision time indicate whether the simulation assumes a step-like ionization profile ($\tau_{ni,0}/t_{0} = 0.001$) or a cosmic ray ionization profile ($\tau_{ni,0}/t_{0} = 0.2$).

\begin{deluxetable}{ccccc}
\tablecaption{Simulation Parameters}
\tablewidth{0pt}
\tablehead{
\colhead{Model} & \colhead{$\mu_{0}$} & \colhead{$\tau_{ni,0}/t_{0}$\tablenotemark{*}} & \colhead{$A$} & \colhead{$\Delta t_{sp}/t_{0}$}
}
\startdata
A & 1.1 & 0.001 & 0.03   & $\infty$  \\ 
B & 1.1 & 0.001 & 0.03   & 5        \\   
C & 1.1 & 0.001 & 0.03   & 10       \\  
D & 1.1 & 0.001 & 0.03   & 15       \\ 
E & 1.1 & 0.001 & 0.015  & 10       \\ 
F & 2.0 & 0.001 & 0.03   & 10       \\  
G & 1.1 &   0.2 & 0.03   & 10        
\enddata
\tablenotetext{*}{The value of $\tau_{ni,0}/t_{0}$ indicates whether the ionization profile was step-like ($\tau_{ni,0}/t_{0}~=~0.001$) or CR only ($\tau_{ni,0}/t_{0}~=~0.2$ ) }
\label{models}
\end{deluxetable}

\section{Simulations and Results}

We ran seven distinct models which differed in the time between perturbations $\Delta t_{sp}/t_{0}$, the amplitude of the perturbations $A$, and the initial conditions of the molecular cloud environment (see Table~\ref{models}). Analysis of all seven models showed that the simulations that vary the value of $\Delta t_{sp}/t_{0}$ (Models B, C, D, and E) result in similar column density and mass-to-flux ratio structures, with small variations due to the difference between these four models. Conversely, Models A (single initial perturbation), F (significantly supercritical $\mu_{0}$), and G (CR ionization only) exhibit significantly different column density and mass-to-flux ratio maps when compared to the other four. We focus mainly on the four models that show significant differences (A, C, F, and G) and include discussion on the variations observed in the other three (B, D, and E) where appropriate. We chose Model C as the representative case among Models B, C, D, and E since it has the median value for $\Delta t_{sp}/t_{0}$.

\begin{figure*}
\includegraphics[width=0.5\textwidth]{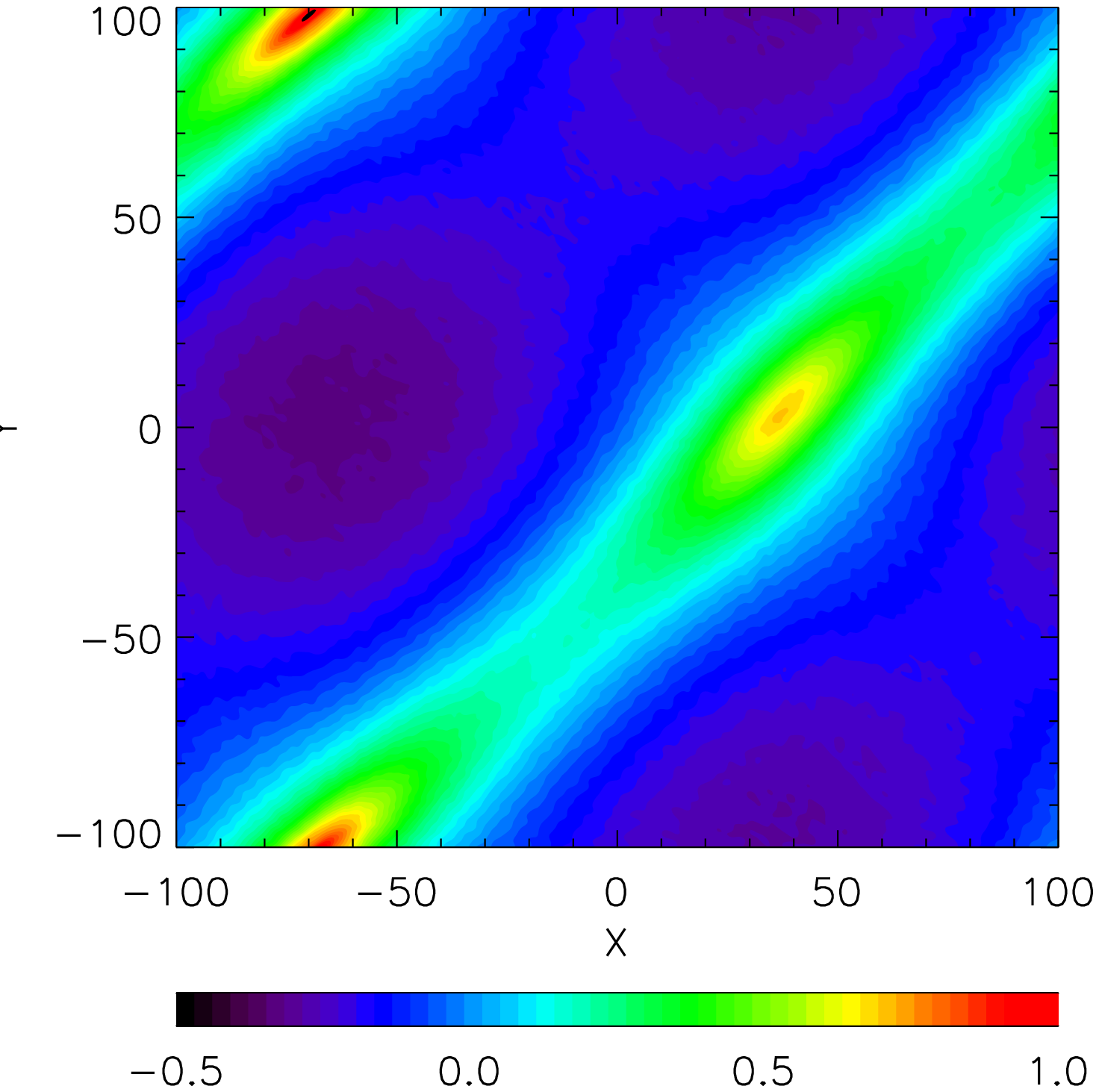}
\includegraphics[width=0.5\textwidth]{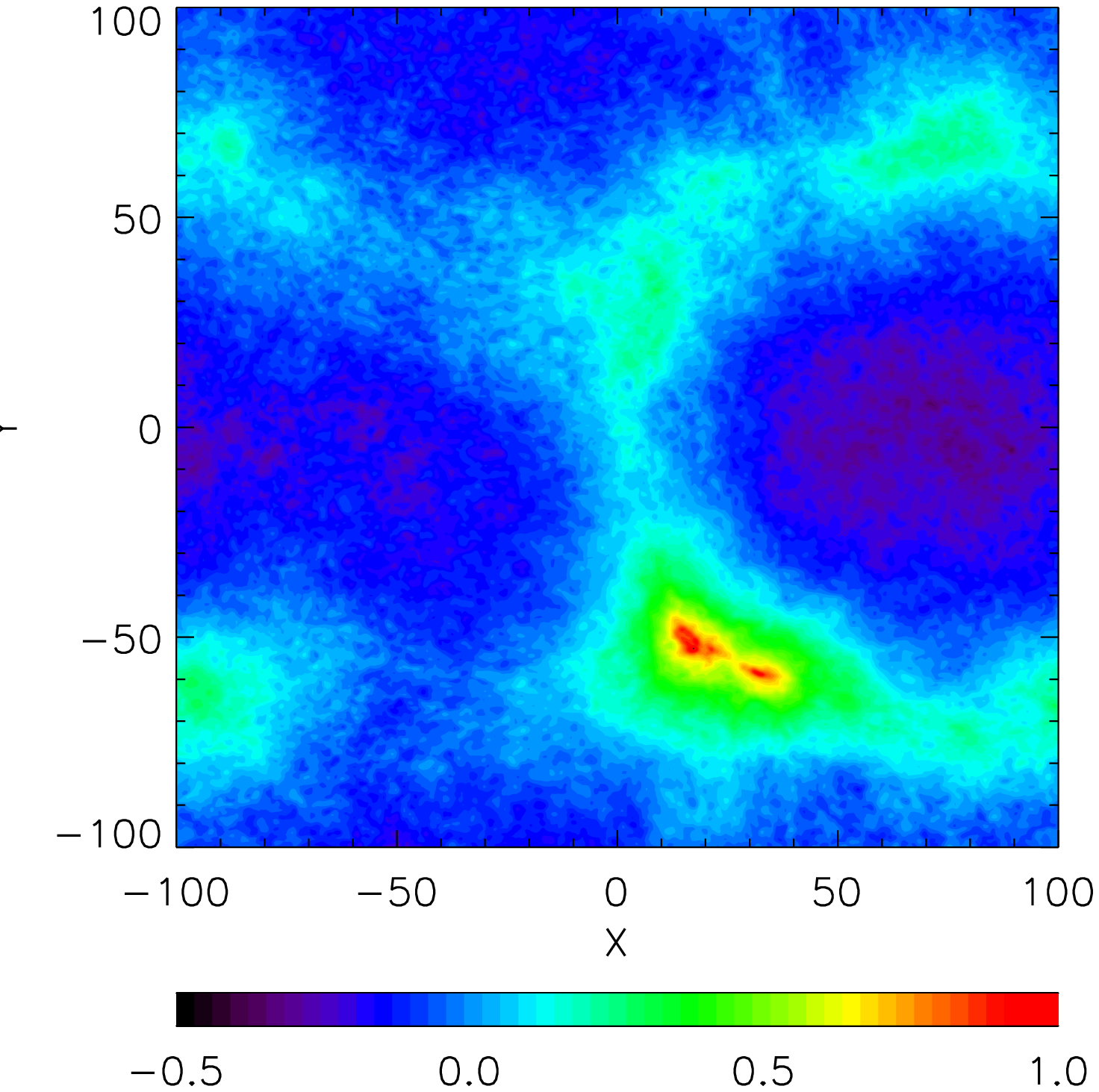}\\
\includegraphics[width=0.5\textwidth]{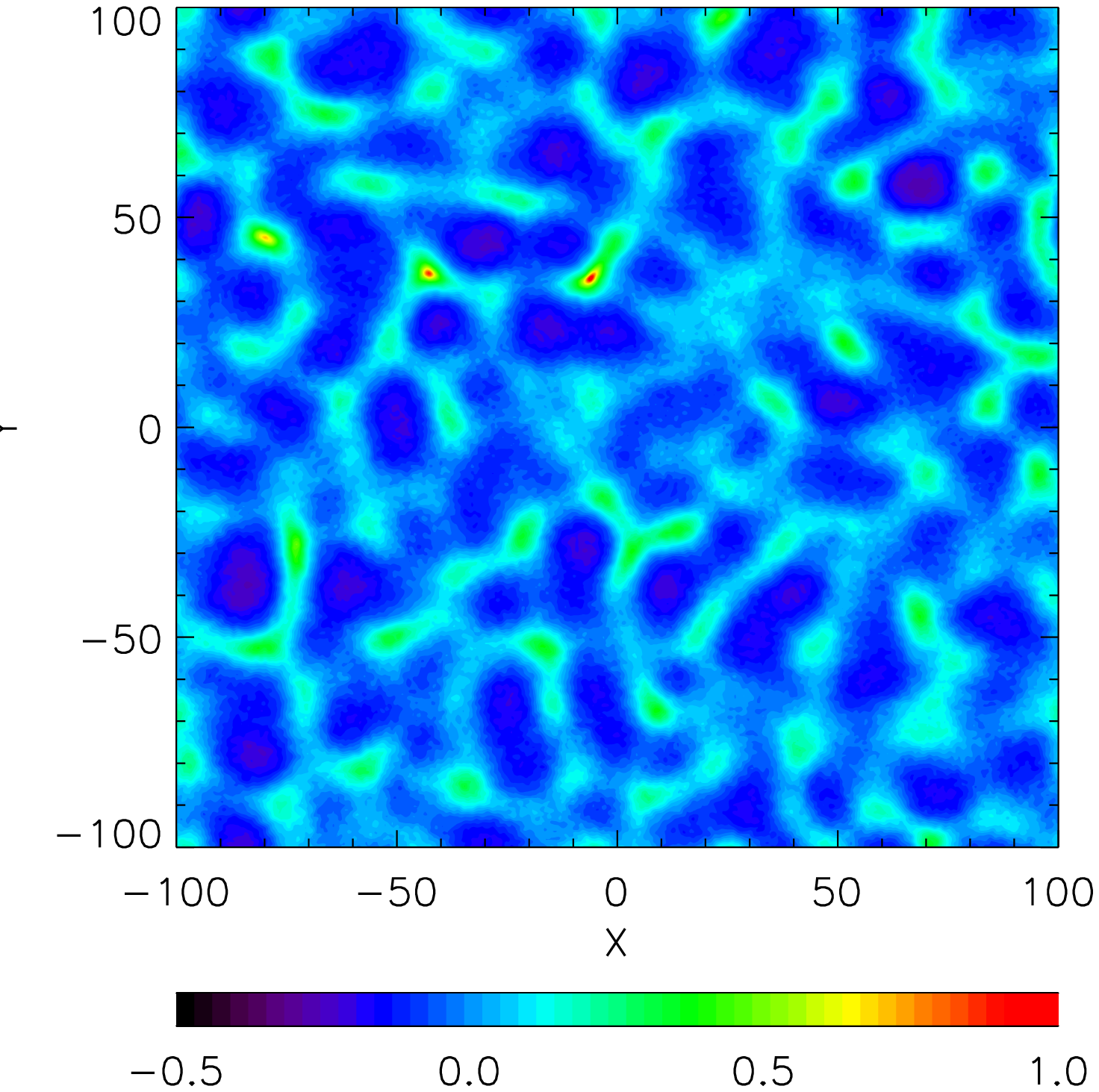}
\includegraphics[width=0.5\textwidth]{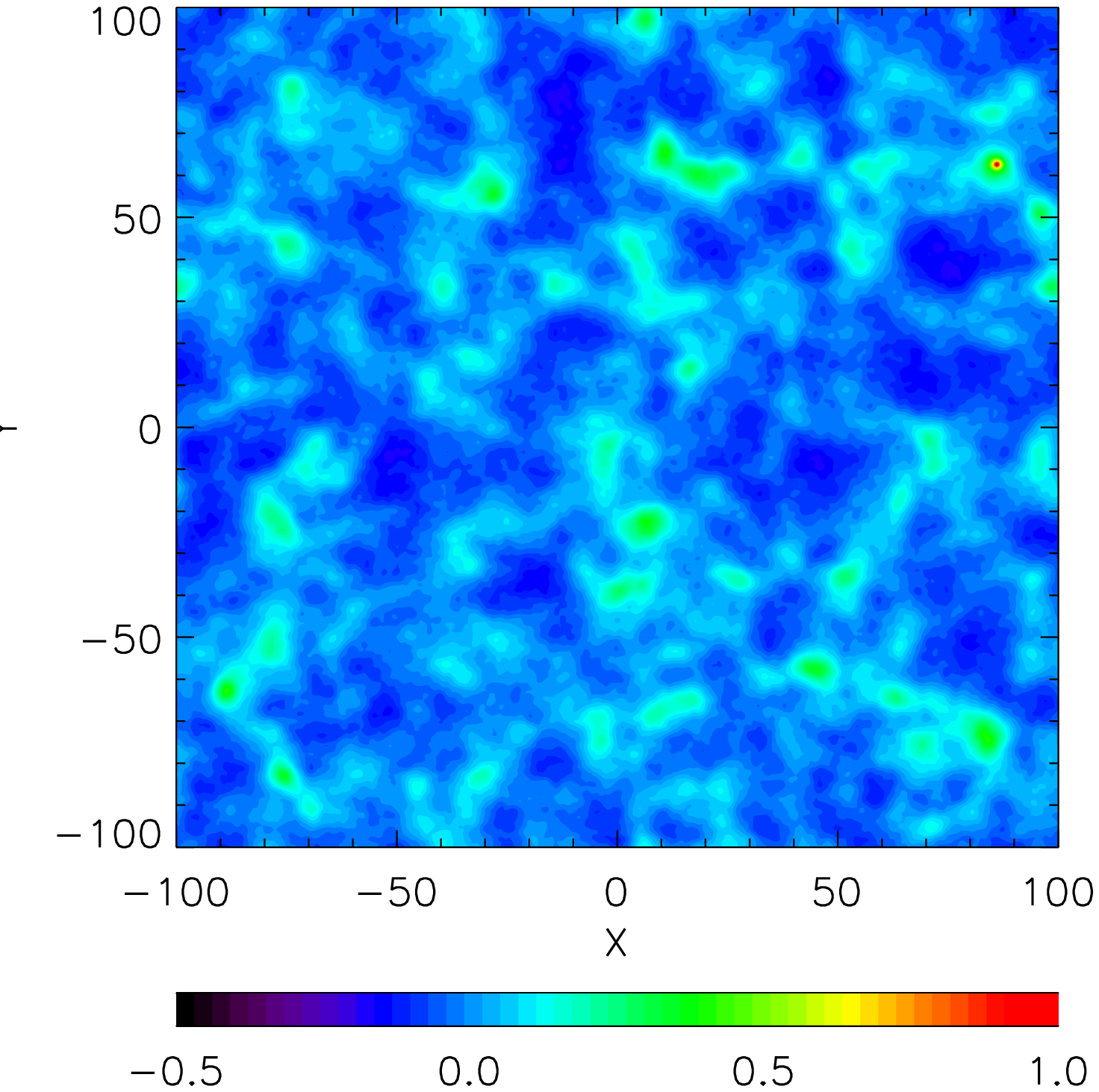}
\caption{Column density enhancement ($\sigma_{n}/\sigma_{n,0}$) maps for Model A (upper left), Model C (upper right), Model F (lower left), and Model G (lower right) at the final times ($t_{\rm run}$) for each model (see Table~\ref{times}). Color bars give the range of column density enhancement values on a $\log_{10}$ scale. Axes give the length in units of $L_{0}$. }
\label{fig:Av}
\end{figure*}

The simulations follow the evolution of several different physical parameters including the column density enhancement ($\sigma_{n}/\sigma_{n,0}$), changes in the $x$- and $y$- momentum, magnetic field strength ($B$), and ionization fraction ($\chi_{i}$). We found that in general, the magnetic field and ionization fraction structures followed that of the column density. However, the mass-to-flux ratio showed differing behaviour depending on the parameters of the simulation. The following sections present some of the results and trends from our simulations.

\subsection{Column Density}

Figure~\ref{fig:Av} shows the column density enhancement maps ($\sigma_{n}/\sigma_{n,0}$) for Model A (upper left), Model C (upper right), Model F (lower left) and Model G (lower right) at the final time for each model. These maps show that the parameters of the model can have a significant effect on structure formation. For Model A (upper left panel), the simulation region fragments into two distinct clumps (due to the periodic boundary conditions, the dense regions in the top left and bottom left of the simulation represent one continuous clump). \textit{There is no evidence of a second fragmentation event}. Rather, the clump exhibits an onion like structure with a high column density central region and layers of more diffuse material as the radial distance from the center increases. 

The upper right panel of Figure~\ref{fig:Av} shows the column density enhancement map for Model C. The addition of perturbations over the course of the simulation results in a very different column density structure than in Model A. There are two closely related high column density clumps in the lower central portion of the simulation region with four other less dense clumps in the upper left, lower left and upper right. Within the two high column density clumps, there is evidence of five distinct substructures that were formed after the ionization fraction within the clump decreased to cosmic ray levels. \textit{Hence, there is a second-stage fragmentation event in this model}. Models B, D, and E (not shown) exhibit similar column density maps with small variations. For Models B and D, the change in time between perturbations affects the shape of the high column density clump and the number of cores formed within. A decrease in the time between perturbation events (Model B) results in a more fragmented clump region that exhibits more cores while an increase in the time between perturbations (Model D) results in a more cohesive clump region that exhibits fewer cores. Reducing the amplitude of the perturbations (Model E) also results in a more cohesive clump with fewer cores. The lower left panel of Figure~\ref{fig:Av} shows the column density enhancement map for Model F. The greater initial mass-to-flux ratio ($\mu_{0} = 2$) of this model causes fragmentation into many smaller clumps with no evidence of subfragmentation. This is consistent with the preferred fragmentation scale $\lambda_{g,m}$ being much smaller for $\mu_{0} = 2$ than for $\mu_{0} = 1.1$ \citep{CB2006, BB2012}. Finally, the lower right panel shows the column density enhancement maps for Model G, which has only cosmic ray ionization. This model, like Model F, fragments into small clumps with no evidence of subfragmentation. In this case, the smaller fragmentation scale is due to the longer neutral-ion collision time.

\begin{figure}
\includegraphics[width=0.4\textwidth, angle=-90]{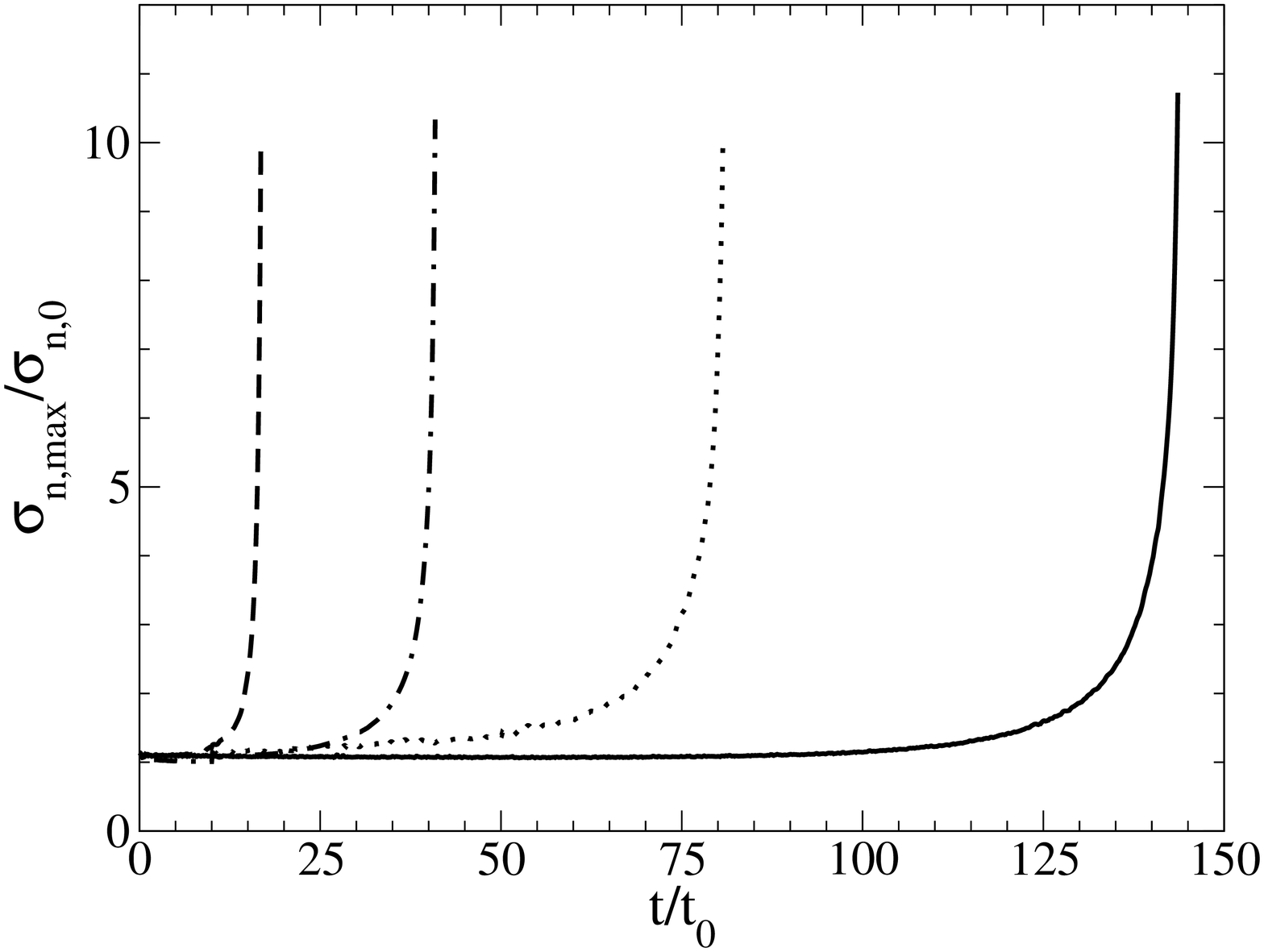}\\
\includegraphics[width=0.4\textwidth, angle=-90]{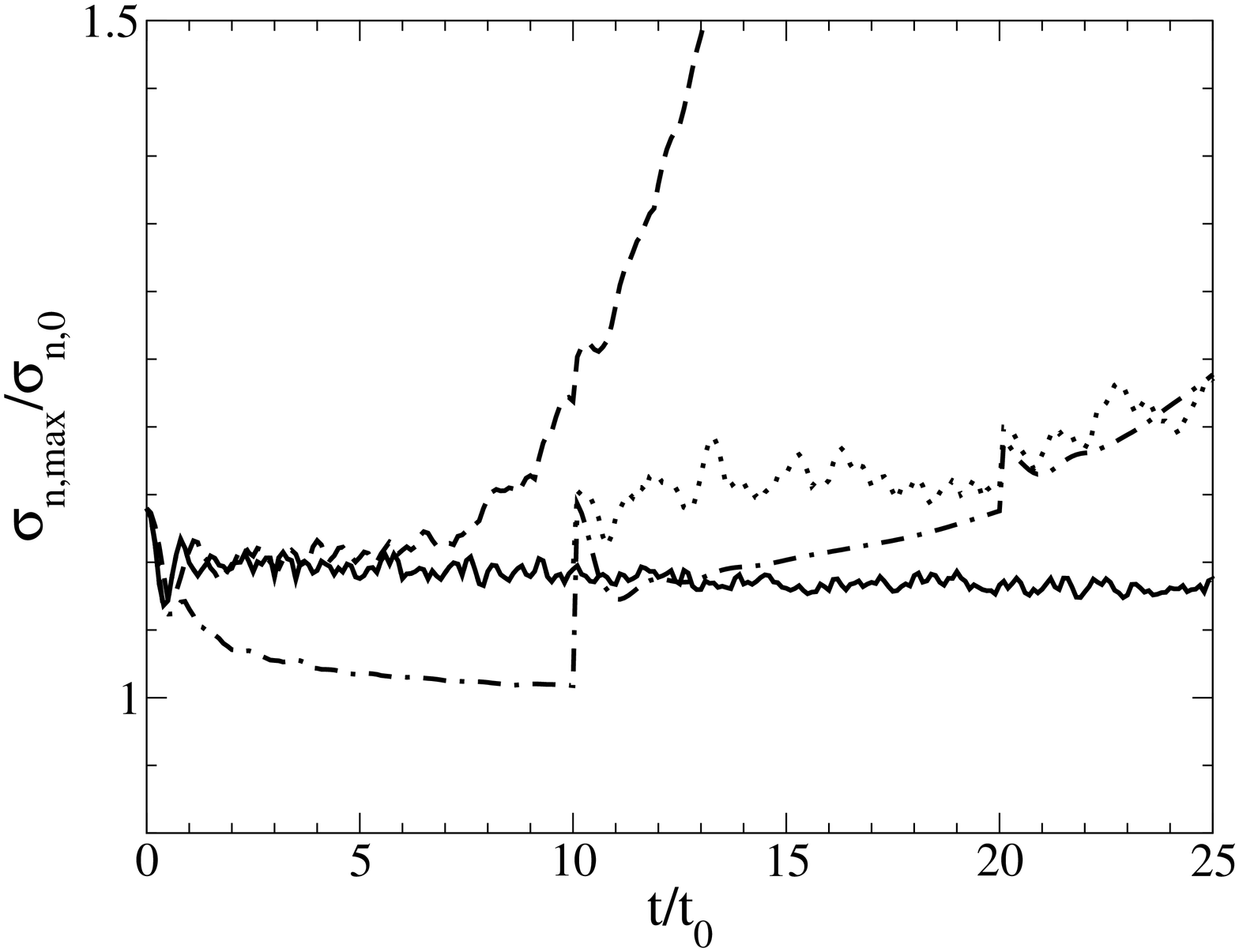}
\caption{Maximum column density enhancement value ($\sigma_{n, \rm max}/\sigma_{n,0}$) within the simulation region as a function of time for Model A (solid line), Model C (dotted line), Model F (dashed line), and Model G (dot-dashed line). Top panel: the full evolution of $\sigma_{n,\rm max}/\sigma_{n,0}$ for all simulations. Bottom panel: variation of $\sigma_{n,\rm max}/\sigma_{n,0}$ at early times. }
\label{fig:max_sigma}
\end{figure}

Figure~\ref{fig:max_sigma} shows the evolution of the maximum value of $\sigma_{n}/\sigma_{n,0}$ ($\sigma_{n,\rm max}/\sigma_{n,0}$) as a function of time for Model A (solid line), Model C (dotted line), Model F (dashed line) and Model G (dot-dashed line). The upper panel shows the evolution over the full simulation time. For each model, $\sigma_{n,\rm max}/\sigma_{n,0}$ increases gradually until it reaches a value between 2 and 3. At this point, the column density increases rapidly over a very short time frame as runaway collapse occurs. The lower panel of Figure~\ref{fig:max_sigma} shows only the early times for the same four models. As shown, all models start with the same value of $\sigma_{n,\rm max}/\sigma_{n,0}$. We see that there are three different evolutionary paths, depending on the model parameters. The first of these paths is defined by Models A and C, for which the instantaneous value of $\sigma_{n,\rm max}/\sigma_{n,0}$ shows fluctuations within an overall increasing trend. Both models follow the same line until Model C reaches the time of its next perturbation. At this point, $\sigma_{n,\rm max}/\sigma_{n,0}$ of Model C jumps abruptly. After each perturbation, the fluctuations of $\sigma_{n,\rm max}/\sigma_{n,0}$ increase. This continues until $\sigma_{n,\rm max}/\sigma_{n,0}$ reaches between 2 and 3, after which it increases dramatically as discussed before. The second path is defined by Model F. Here, the increased value of $\mu_{0}$ results in larger fluctuations of $\sigma_{n,\rm max}/\sigma_{n,0}$ at early times. This allows the simulation to reach a point of runaway collapse faster than the other simulations, specifically Model C. The final path is that of Model G. Here, after the initial perturbation, $\sigma_{n,\rm max}/\sigma_{n,0}$ drops dramatically compared to the other models.  There are also less fluctuations overall. After each perturbation event, $\sigma_{n,\rm max}/\sigma_{n,0}$ continues to exhibit an initial drop, however the extent of the decrease is smaller each time and the overall trend is increasing. The sudden decrease after the perturbation event is a result of an internal pressure gradient pushing the neutral particles back across the magnetic field lines. This phenomenon is evident in the other models, however the lower ionization fraction of this model allows for neutrals to move more freely resulting in a more significant decrease.

\begin{table*}
\centering
\caption{Model timescales}
\begin{tabular}{ccccccccccccc}
\hline\hline
&& \multicolumn{2}{c}{$t_{\rm run}$} && \multicolumn{2}{c}{$t_{\rm frag}$} && \multicolumn{2}{c}{$t_{\rm clump}$} && \multicolumn{2}{c}{$t_{\rm core}$}\\
\cline{3-13}
Model && $t/t_{0}$ & Myr\tablenotemark{a} && $t/t_{0}$ & Myr\tablenotemark{a} && $t/t_{0}$ & Myr\tablenotemark{a} && $t/t_{0}$ & Myr\tablenotemark{a} \\
\hline
A  && 143.6 & 56.3 && 40 - 50 & 15.6 -19.6 && 135      & 52.9        && \nodata  & \nodata  \\
B  && 70.0  & 27.5 && $\sim$10& $\sim$3.9  && $\sim$60 & $\sim$23.5  && 65.1     & 25.5\\
C  && 80.7  & 31.7 && 10 - 20 & 3.9 - 7.8  && $\sim$70 & 27.5        && 75       & 29.4\\
D  && 87.7  & 34.4 && 10 - 20 & 3.9 - 7.8  && 75 - 80  & 29.4 - 31.4 && $\sim$80 & 31.4\\
E  && 95.2  & 37.3 && 10 - 20 & 3.9 - 7.8  && 80 - 85  & 31.4 - 33.3 && $\sim$85 & 33.3\\
F  && 16.8  &  6.6 && $<$10   & $<$3.9     && $\sim$15 & $\sim$5.9   && \nodata  & \nodata  \\
G  && 40.9  & 16.0 && $<$10   & $<$3.9     && $\sim$20 & $\sim$7.8   && \nodata  & \nodata\\
\hline
\label{times}
\end{tabular}
\tablenotetext{1}{Time scales in Myr determined by multiplying previous column by $t_{0}$ assuming T = 10 K and $\sigma_{n,0} = 3.638 \times 10^{-3}$ g cm$^{-2}$.} 
\end{table*}

Figure~\ref{fig:max_sigma} also reveals the length of each evolutionary stage. There are four evolutionary stages that are of particular interest: the total run time ($t_{\rm run}$), the initial large scale fragmentation time of the cloud ($t_{\rm frag}$, i.e., the time when large scale fragmentation starts to occur with $\sigma_{n}/\sigma_{n,0}$ still close to unity), the clump formation time ($t_{\rm clump}$), defined as the time when the outer boundaries of the clump region reaches $\sigma_{n}/\sigma_{n,0} > 2$, and the core formation time ($t_{\rm core}$), defined as the time when distinct substructures form within clumps. Table~\ref{times} shows these four evolutionary times for all seven models. A range of ages indicate that the event occurred at some point within that period, however the exact time is not known due to the frequency of data output. Comparing Models A, B, C, and D, we see that the total run time of the simulation decreases as the frequency of perturbations increases. With ongoing perturbations, the overall simulation time is shortened by approximately a factor 2 for Model B, and by a factor 1.6 for Model D. Models E, F, and G see varying effects on $t_{\rm run}$. The decrease in perturbation amplitude in Model E results in an increase in total run time of about 19\% compared to Model C. Conversely, the increase of $\mu_{0}$ in Model F and initially lower $\tau_{ni}/t_{0}$ value in Model G have the opposite effect, reducing the run time by factors of 4 and 1.75, respectively. Table~\ref{times} also shows that the initial fragmentation within the region typically occurs very early in the simulation for each model. This process takes up about $10\% - 20\%$ of the simulation time. The majority of the simulation time ($\sim 65\%$) goes toward forming the clump while the final $\sim 5\%$ of the simulation is the formation and collapse of the core region. Further comparison to the two-stage fragmentation model will be presented in Section~\ref{simdiss}.

Looking at the times for each stage quantitatively, the age of the system at the end of the simulations is between 27 and 37 Myr for Models B, C, D and E while Models A, F and G have ages of 56.3, 6.6, and 16 Myr, respectively. These time frames are of particular interest when considering the total lifetime of a molecular cloud. Observational and theoretical estimates of molecular cloud lifetimes fall within two categories: short (less than a few Myr) \citep{Hartmann2001, HBB2001} and long (greater than 20~-~30 Myr) \citep[][among others]{BS1980,XLB1984, Murray2011}. The main argument for short lifetimes hinges on the ages of the young stellar objects (YSOs) observed within the molecular cloud. YSOs are typically quoted to have ages on the order of a few Myr. In addition, the ionization from newly formed high mass stars, where present, may disrupt the molecular cloud a few Myr after their formation. This line of reasoning only considers the age of the YSO itself as a proxy for the age of the molecular cloud. It does not include the time that it takes for the molecular cloud to collapse and form the pre-stellar core and YSO. The typical age of our simulation cloud at the formation of the first core (i.e., a pixel reaches $\sigma_{n}/\sigma_{n,0} = 10$) is greater than 27 Myr. This is consistent with long lifetimes of molecular clouds \citep[see][]{Murray2011}, with most of the lifetime occupied with clump formation, followed by a relatively rapid phase of core and star formation. The two cases with shorter lifetimes (Models F and G) represent systems in which the evolution is already advanced, i.e., the cloud already has a region with a supercritical mass-to-flux ratio or a smaller ionization fraction. Given that most molecular clouds do not form with these conditions, but rather must evolve to that state, these two simulations can be considered to be only a portion of the full evolution.  

\begin{figure}
\includegraphics[width=0.5\textwidth,angle=-90]{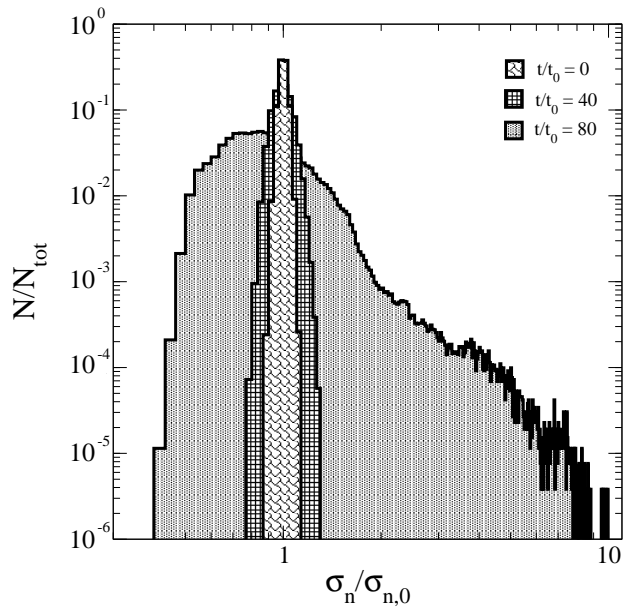}
\caption{Histogram of the column density enhancement for Model C. Shaded regions show the distributions for three different times as indicated.}
\label{fig:hist}
\end{figure}

Figure~\ref{fig:hist} shows a histogram of the column density enhancement for Model C. Shaded regions show the distribution of column density at three different times, $t/t_{0}$ = 0.1, $t/t_{0}$ = 40, and $t/t_{0}$ = 80, as indicated on the plot. As shown, the initial distribution is a narrow Gaussian centered on $\sigma_{n}/\sigma_{n,0} = 1$. The width of this initial Gaussian in indicative of the amplitude of the perturbation applied at the beginning of the simulation. As the simulation evolves, we see that the width of the Gaussian increases as some regions within the simulation accumulate mass while others contemporaneously lose mass. At the end of the simulation ($t/t_{0}$ = 80), the histogram shows a Gaussian peak with a high column density power law tail. These two different distributions (pure Gaussian and Gaussian with power-law tail) have been observed within molecular clouds by \citet{Kain2009}. In these observations, the pure Gaussian profiles were associated with quiescent non-star-forming clouds, while power-law tails were observed in regions with active star formation. The histogram in Figure~\ref{fig:hist} shows that the tail develops sometime after $t/t_{0} = 40$ or 16 Myr, indicating that molecular cloud lifetimes must be on the order of 20~-~30 Myr based on our models. Similar distributions were also found in highly turbulent simulations by \citet{Cho2011}, although it is not clear how to trace the evolution time in their model back to the origin of the molecular cloud.
\begin{figure*}
\includegraphics[width=0.5\textwidth]{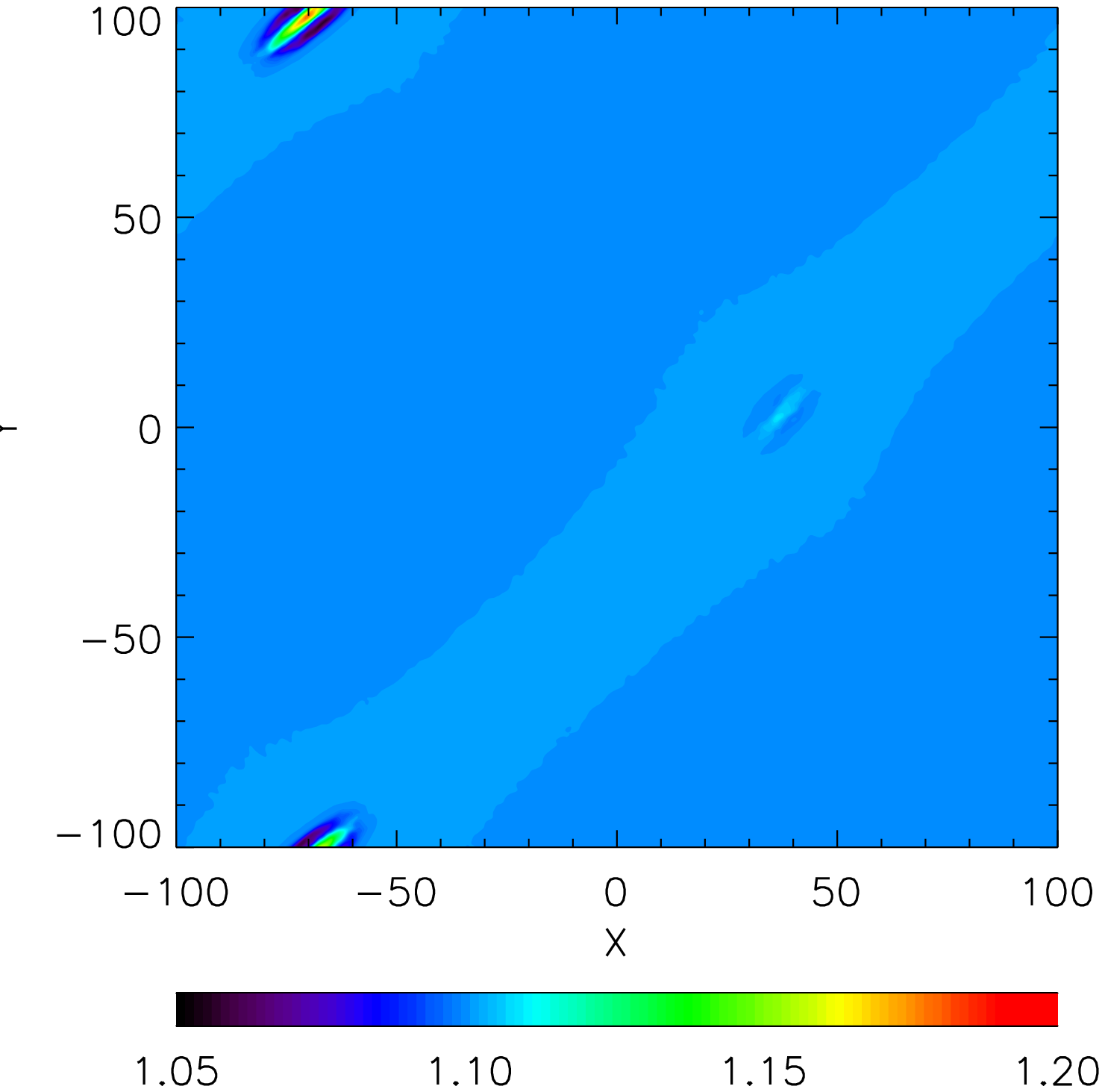}
\includegraphics[width=0.5\textwidth]{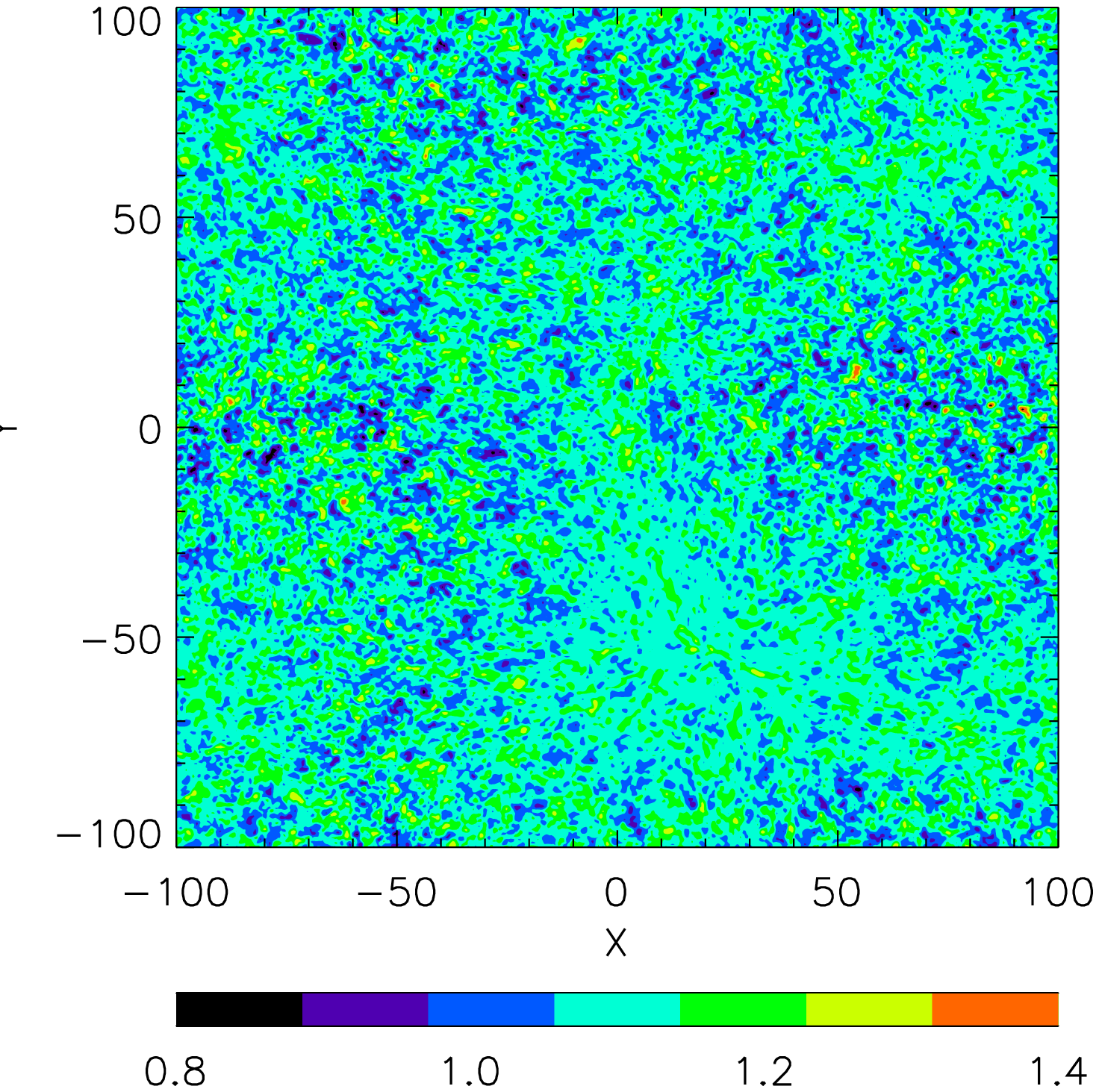}\\
\includegraphics[width=0.5\textwidth]{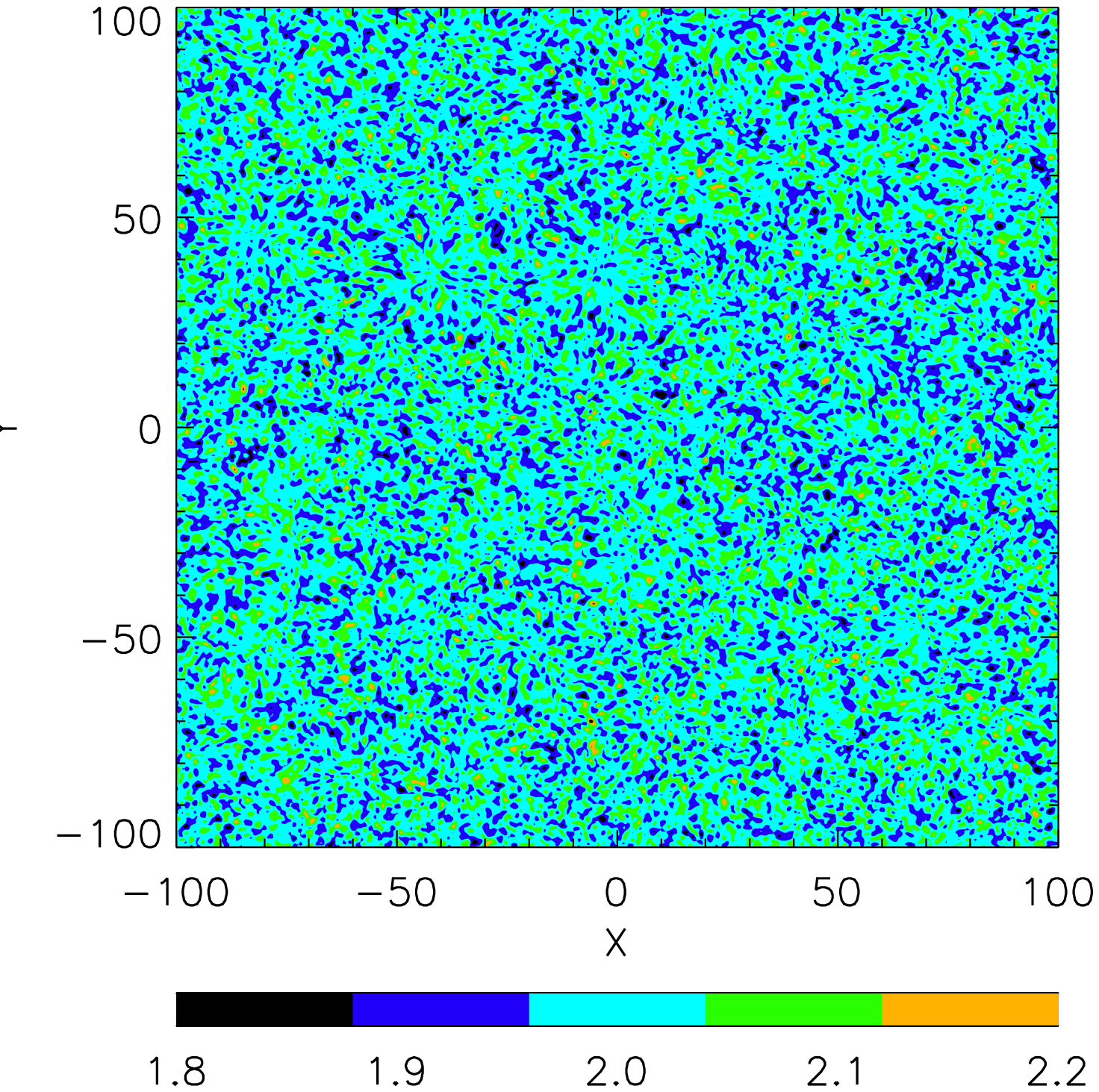}
\includegraphics[width=0.5\textwidth]{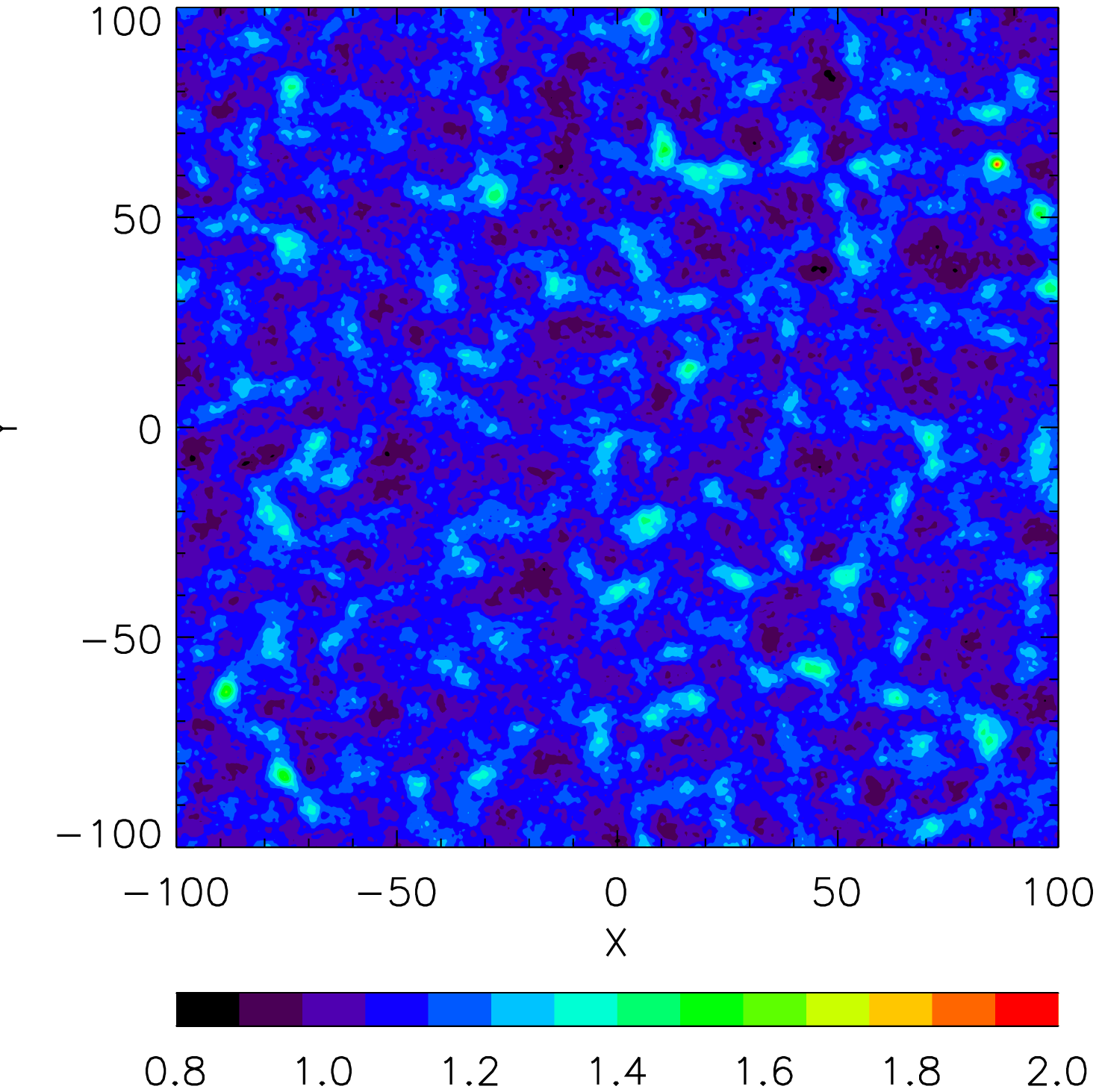}
\caption{Mass-to-flux ratio ($\mu$) maps for Model A (upper left), Model C (upper right), Model F (lower left), and Model G (lower right) at the final times ($t_{\rm run}$) for each model (see Table~\ref{times}). Color bars give the range of $\mu$ values on a linear scale. Axes give the length in units of $L_{0}$. }
\label{fig:mtf}
\end{figure*}

\subsection{Mass-to-flux ratio}

Figure~\ref{fig:mtf} shows the mass-to-flux ratio maps for Model A (upper left), Model C (upper right), Model F (lower left), and Model G (lower right) at the final time for each model. For Model A, $\mu$ tends to follow the column density contours (see Figure~\ref{fig:Av}, upper left), while the addition of multiple perturbations during the evolution of the cloud results in a more randomized field of values. Looking closer at the mass-to-flux ratio map for Model A (Figure~\ref{fig:mtf}, top left), the clump region exhibits a peak in mass-to-flux ratio that is surrounded by a region of low mass-to-flux ratio that is less than the background value. For Models C and F (Figure~\ref{fig:mtf}, top right and bottom left respectively), we see that the mass-to-flux ratio field is much more chaotic. Some coherent structure forms around the high column density regions, however it is hard to pick out a particular clump region within these mass-to-flux ratio maps unless the location is already known. Finally for Model G (Figure~\ref{fig:mtf}, bottom right), the mass-to-flux ratio field again follows the column density map. This is a consequence of the CR only ionization profile in this model. This implies that the irregular nature of the mass-to-flux ratio maps for Models C and F are \textit{due to the combination of multiple perturbations during the simulation time and a high ionization fraction for the majority of that time}. Under these circumstances, ambipolar diffusion is not efficient enough to readjust the matter to attain a smooth mass-to-flux ratio distribution as exhibited by Model A.

\begin{figure}
\includegraphics[width=0.4\textwidth, angle=-90]{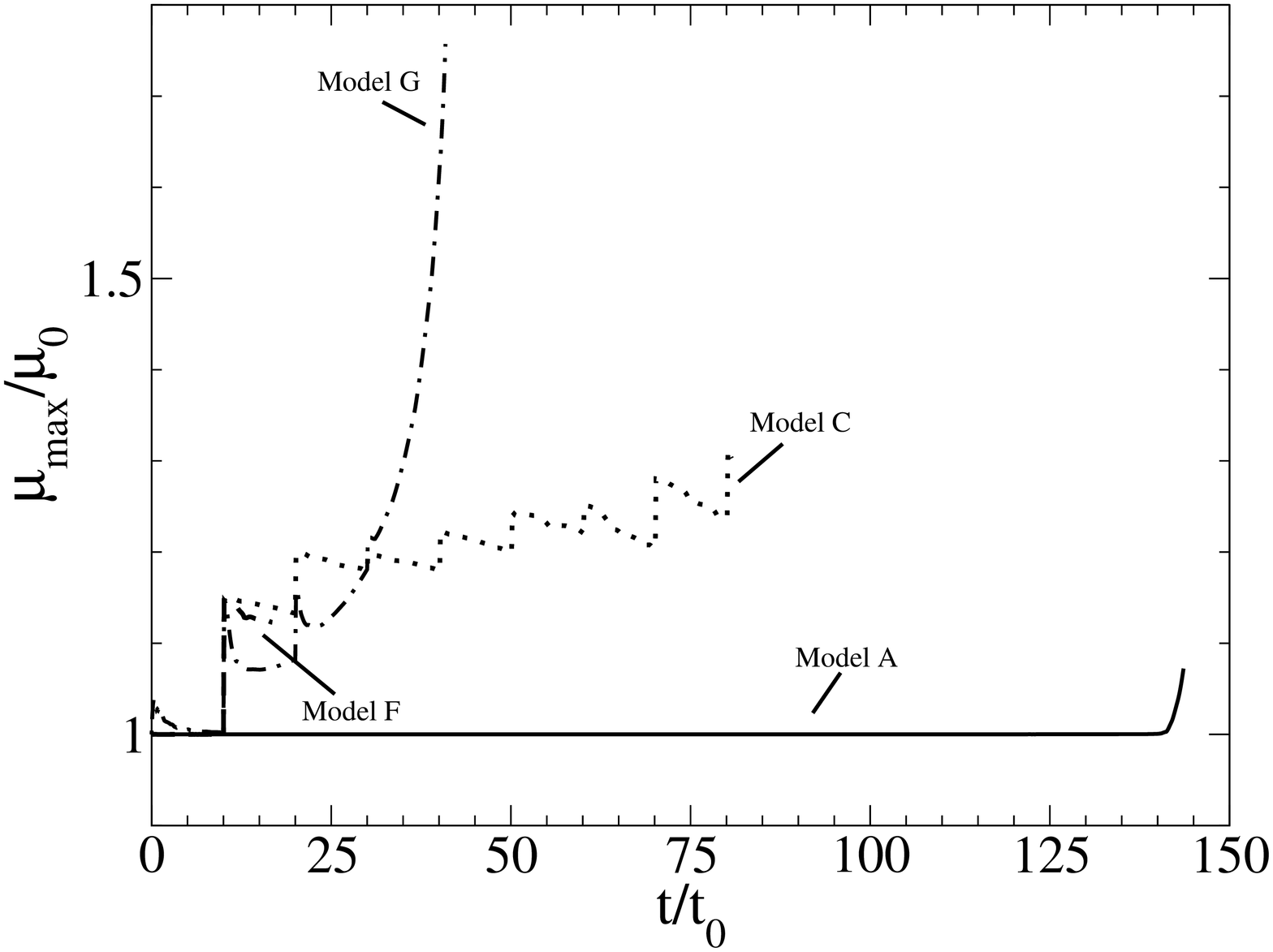}
\caption{Scaled maximum mass-to-flux ratio ($\mu_{\rm max}/\mu_{0}$) within the simulation region as a function of time for Model A (solid line), Model C (dotted line), Model F (dashed line), and Model G (dot-dashed line). Scaling is based upon the initial mass-to-flux ratio for each model. }
\label{fig:max_mtf}
\end{figure}

Figure~\ref{fig:max_mtf} shows the evolution of the scaled maximum mass-to-flux ratio ($\mu_{\rm max}/\mu_{0}$) as a function of time for Models A, C, F, and G. Note that the location of this maximum may not be depicted by the same pixel at all times. The evolution of $\mu_{\rm max}/\mu_{0}$ in Model A is similar to the evolution of $\sigma_{n,\rm max}/\sigma_{0}$ depicted in Figure~\ref{fig:max_sigma}, which confirms that without additional ongoing perturbations, $\mu_{\rm max}/\mu_{0}$ coincides with the location of the column density peak. The additional perturbations present in Models C and F results in a time evolution with a saw-tooth like pattern. Finally, Model G exhibits a much different evolution  for $\mu_{\rm max}/\mu_{0}$. Recall that this simulation assumes a CR only ionization profile with $\tau_{ni,0}/t_{0} = 0.2$. This model exhibits the greatest mass-to-flux ratio increase out of all the models presented. This is again a direct consequence of the ionization profile and longer neutral-ion collision time within this model. Due to increased neutral-ion slip, the column density can increase while the magnetic field strength stays relatively constant, resulting in a significant increase to the mass-to-flux ratio. In the other models, the ionization fraction is much greater at early times, resulting in an almost flux-frozen medium. Initial redistribution of mass within these simulations would also drag the magnetic field with it, thus maintaining a mass-to-flux ratio near the initial value.

\subsection{Clumps and Cores}

\begin{figure}
\includegraphics[width=0.5\textwidth]{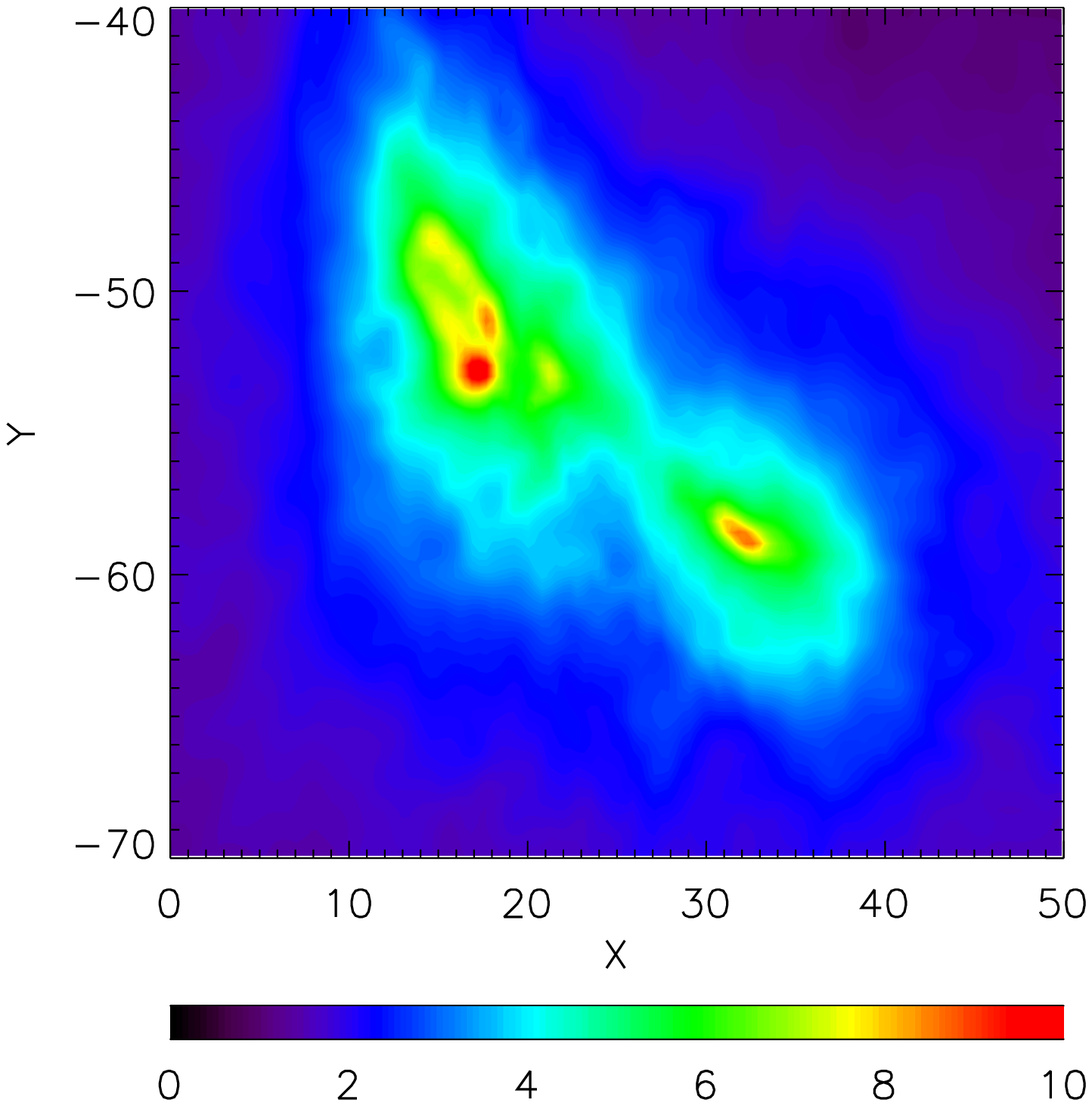}\\
\includegraphics[width=0.5\textwidth]{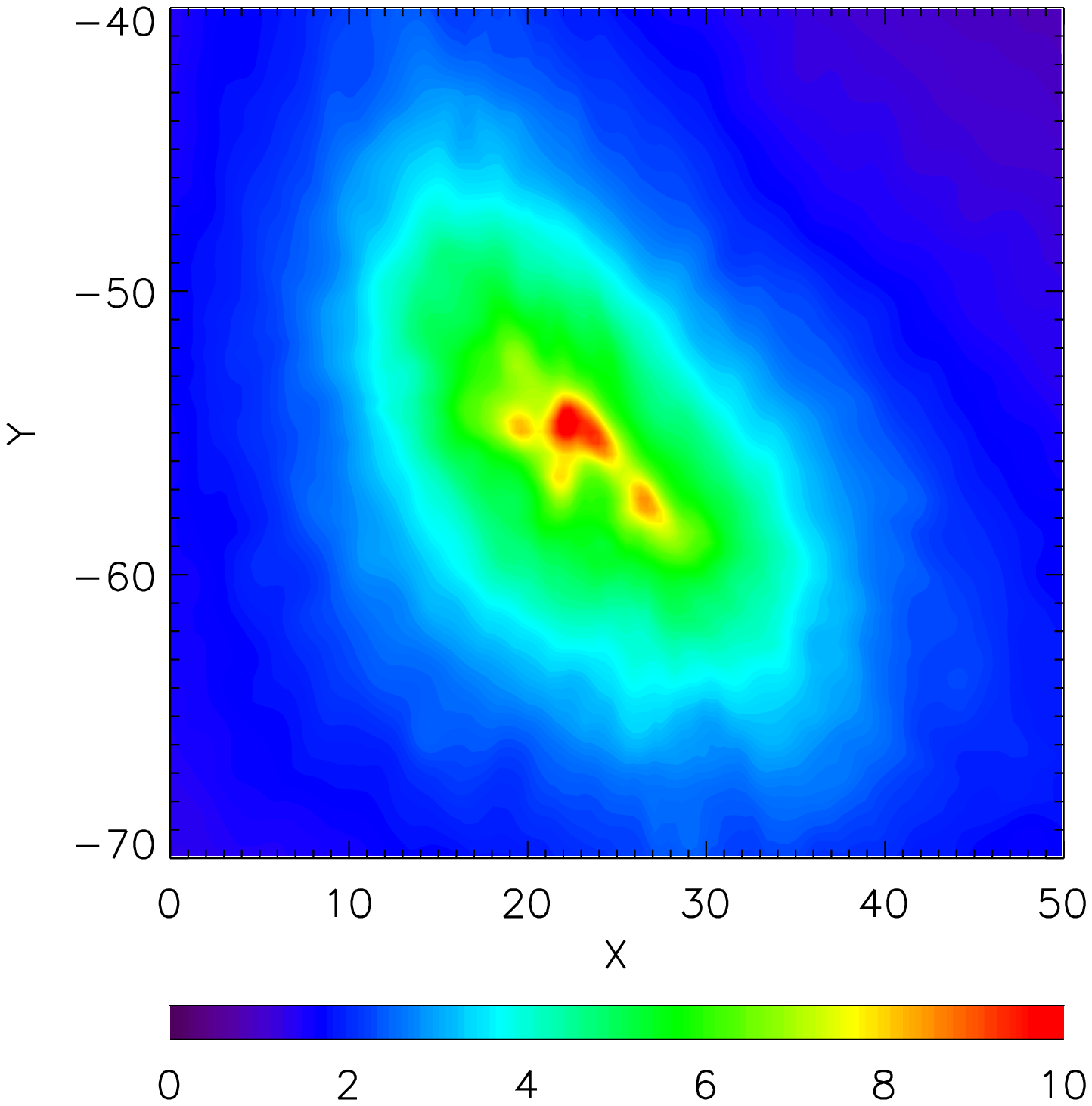}
\caption{Column density enhancement ($\sigma_{n}/\sigma_{n,0}$) map of a prominent clump region in Model C (left panel) and Model E (right panel) at the final time ($t_{\rm run}$). Color bars give the range of column density enhancement values on a linear scale. Axes give the length in units of $L_{0}$. }
\label{fig:cores}
\end{figure}

One of the main analyses performed on a molecular cloud is the identification of clump and core regions and the determination of their physical properties (e.g., radius, column density, mass, etc.). Here we perform such an analysis on our simulations. Figure~\ref{fig:cores} shows a blown up map of the prominent clump region in Model C (upper) and Model E (lower). For Model C, there are three visible cores, while for Model E, there are 5 visible cores.  Visual inspection of this prominent clump in Models B, C, D, and E shows that for Models B, C, and D, the number of visible cores decreases as the frequency of perturbations decreases. Another trend evident in Models B, C, and D is the decreasing average column density of the other clumps within the simulation region as the frequency of perturbations decreases. This is likely another consequence of the medium having more time to readjust itself after a perturbation event. Looking at the structure of the prominent clump itself, we see that the frequency and amplitude of the perturbations has an effect on its structure. Model B ($\Delta t_{sp} = 5t_{0}$) exhibits three distinct regions, Model C ($\Delta t_{sp} = 10t_{0}$) has two regions, while Model D ($\Delta t_{sp} = 15t_{0}$) has only one. Comparing Models C and E (see Figure~\ref{fig:cores}), we see that the effect of decreasing the amplitude of the perturbation is again a decrease in the amount of structure: Model C shows evidence of two clump regions while only one is evident in Model E. Finally looking at the velocity maps, we see that the magnitude of infall speed within the core is typically subsonic to transonic, in agreement with other numerical models \citep[e.g.,][]{BC2004,Basu2009b} and analytic models \citep{AS2007} of core formation driven by ambipolar diffusion.

In addition to these general trends, we can also quantitatively define clump and core regions and determine their physical properties (e.g., radius and mass). Observationally, the definition of a clump/core depends on the region in which they are being observed, however there are some typical visual extinction thresholds for these different structures. Clump regions are generally defined as coherent regions where the visual extinction is $A_{V} > 1~-~3$ mag. Studies of the Ophiuchus and Perseus clouds by \citet{Johnstone2004} and \citet{Kirk2006}, respectively, suggest that the core formation threshold is  $A_{V} >$ 5 mag and a star formation threshold of $A_{V} \sim 7-8$ mag \citep[see also][]{oni98,Froebrich2010}. 

In general, the threshold column density for the definition of a core is on the order of $N_{H_{2}} = 10^{22}$ cm$^{-2}$. In their analysis of cores produced by the thin-disk code used here, \citet{Basu2009b} defined the background column density to be equal to this threshold, i.e., $N_{n,0} = 1.0\times 10^{22}$ cm$^{-2}$. In order to test the two-stage fragmentation model, our initial column density needed to be lower and therefore the background column density was set to $\sigma_{n,0} = 3.638\times 10^{-3}$ g cm$^{-2}$ ($N_{n,0} = 9.3\times 10^{20}$ cm$^{-2}$), which corresponds to a visual extinction $A_{V} = 1.0$ mag. Therefore, at the end of the simulations, $\sigma_{n}/\sigma_{n,0}$ = 10 corresponds to a column density that is just below the typical threshold definition of a core. For our simulations, we look at the mass enclosed within the regions defined by the typical visual extinction thresholds for clumps and cores as discussed above. Specifically, we define a clump region to be that defined by the contour where the column density is a factor 2 above the background and a core region to be those defined by a contour where the column density is $\approx$ 8 times the background.

\subsubsection{Method}

To determine the outer boundaries of a structure of interest within a simulation we use {\sc clumpfind2d} \citep{Williams1994}. Briefly described, this is a set of IDL routines which first determines the extent of structures within observations or simulated data and then analyzes the identified clump/core regions. The routines assume linearly spaced contours based upon user definitions and traces structures by connecting pixels, at each contour level, that are within one resolution element of each other \citep{Williams1994}. For our simulations, we use {\sc clfind} and {\sc clstat} to find the regions with column density enhancements above our defined clump and core thresholds ($A_{V} = 2$ mag and 8 mag, respectively). Output from these routines include the ``intensity'' of the pixels within the identified clump/core and the effective circular radius as defined in \citet[][Equation A3]{Williams1994}. In our case, the intensity is the sum of the column density over all pixels within the defined clump/core. From this data, we then determine the physical radii in parsecs and enclosed mass within these regions.

\subsubsection{Results}

We perform clump/core analysis on the column density enhancement data for all seven models. For consistency, we perform this analysis at the final time for each of the simulations. 
Table~\ref{clumpparams} shows the derived values for the clump regions found by {\sc clumpfind2d}. The effective radius ($R_{\rm eff}$) is determined by multiplying the radius output from {\sc clumpfind2d} by the dimensional size of a pixel in parsecs  (1 pixel = 0.0296 pc). The mass is calculated by multiplying the total column density found by {\sc clfind} by the dimensional area of a pixel. Comparing all clumps analysed, we see that the radii range from 0.23 pc - 1.61 pc while the masses range from 5.19 M$_{\odot}$ - 412 M$_{\odot}$. Comparing the derived masses, we see that for models B, C, and D, the mass contained within the clumps generally decreases as the frequency of perturbations increases. In addition, as shown by Model E, the mass contained within the clumps increases when the amplitude of the perturbations decreases. We also note that the number of distinct clumps also decreases as the frequency of perturbations increases.

\begin{table}
\centering
\caption{Clump parameters}
\begin{tabular}{ccc}
\hline\hline
Clump \# & ($R_{\rm eff}$) &   Clump Mass ($M_{\rm cl}$) \\
        &      (pc)       &        ($M_{\odot}$)    \\
\hline\hline
\multicolumn{3}{c}{Model A}\\
\hline
A I & 1.61 &  412 \\
\hline
\multicolumn{3}{c}{Model B}\\
\hline
B I   & 0.51 &  47.3\\
B II  & 0.74 &  91.5\\
B III & 0.30 &  11.6\\
B IV  & 0.30 &  11.0\\
B V   & 0.32 &  11.9\\
\hline
\multicolumn{3}{c}{Model C}\\
\hline
C I  & 0.51 & 47.3\\
C II & 0.86 & 122\\
\hline
\multicolumn{3}{c}{Model D}\\
\hline
D I  & 0.82 & 121\\
D II & 0.85 & 113\\
\hline
\multicolumn{3}{c}{Model E}\\
\hline
E I  & 0.90 & 139\\
E II & 0.87 & 133\\
\hline
\multicolumn{3}{c}{Model F}\\
\hline
F I   & 0.27 & 12.1\\
F II  & 0.22 & 8.38\\
F III & 0.23 & 8.08\\
\hline
\multicolumn{3}{c}{Model G}\\
\hline
G I & 0.17 & 5.19\\
\hline\hline
\label{clumpparams}
\end{tabular}
\end{table}

Table~\ref{coreparams} shows the derived values for all of the cores identified by {\sc clumpfind2d} as per the threshold value outlined previously. For each core, we calculate the mass of the region enclosed by the indicated contour. For the densest cores in each model, we also calculate the mass within a smaller region corresponding to a greater contour value, indicated by the ``a'', ``b'', etc. subcategorization (e.g., B1a, C1a, etc). In some cases, although a core region contained pixels with column density enhancement values above the threshold value of 8 mag, the number of pixels was not sufficient enough to be identified as a core by {\sc clumpfind2d}. In these cases, we lowered the value of the contour until these regions contained enough pixels to be found by {\sc clumpfind2d}. All identified cores contain at least 6 pixels. We consider this an acceptable number given that the simulations are at their individual end points.

Comparing all cores analyzed, we see that their sizes are all on the order of 0.1 pc across ($R_{\rm eff} \sim 0.05$ pc) and the masses are in the range ($\sim 0.7 - 3.6$) M$_{\odot}$. The exact values of the radii and mass are directly linked to the contour level used to define the outer boundaries. Defining cores by contours greater than $A_{V} = 8$ mag resulted in smaller core sizes and masses. Comparing to several observational studies \citep[][among others]{Sadavoy2012, Schmalzl2010, Frau2010, RZ2009}, we see that our simulations are producing cores with similar sizes and masses. 

As shown in Table~\ref{coreparams}, the maximum number of cores detected within a single simulation is five, however, this should not be taken as the maximum number that can be produced. Recall that these simulations are stopped after the column density/visual magnitude within any pixel has increased by a factor of ten from the initial state. Further evolution could give rise to other clump-core complexes. Exploration of these later evolutionary stages is left for a future study.

\begin{table}
\centering
\caption{Core parameters}
\begin{tabular}{cccccc}
\hline\hline
Core \# &          Peak Value          & Contour & ($R_{\rm eff}$) &  Core Mass ($M_{\rm c}$) \\
        & ($\sigma_{n}/\sigma_{n,0}$)  & ($\sigma_{n}/\sigma_{n,0}$) &     (pc)       & ($M_{\odot}$)    \\
\hline\hline
\multicolumn{5}{c}{Model B}\\
\hline
B1  & 10.29 & 8.0& 0.055 &  1.57\\
B1a & 10.29 & 9.0& 0.044 &  1.04\\
B2  & 9.15  & 8.0& 0.041 &  0.80\\
\hline
\multicolumn{5}{c}{Model C}\\
\hline
C1  & 10.09 & 8.0  & 0.067 &  2.17\\
C1a & 10.09 & 9.0  & 0.041 & 0.89\\
C2  & 8.65  & 7.92 & 0.041 & 0.75 \\
C3  & 8.55  & 7.92 & 0.041 & 0.75 \\
\hline
\multicolumn{5}{c}{Model D}\\
\hline
D1  & 10.56 & 8.0  & 0.085 &  3.58\\
D1a & 10.56 & 9.0  & 0.058 &  1.76 \\ 
D1b & 10.56 & 9.5  & 0.047 &  1.20 \\
D1c & 10.56 & 9.56 & 0.041 &  0.91 \\
D2  & 8.37  & 8.0  & 0.047 &  1.00 \\
\hline
\multicolumn{5}{c}{Model E}\\
\hline
E1  & 10.44 & 8.0  & 0.077&  2.91\\
E1a & 10.44 & 9.66 & 0.041&  0.92\\
E2  &  9.45 & 8.0  & 0.065&  2.01\\
E2a &  9.45 & 9.0  & 0.041&  0.84 \\
E3  &  8.55 & 8.0  & 0.041&  0.77\\
E4  &  8.25 & 7.6  & 0.050&  1.10\\
E5  &  7.94 & 7.6  & 0.041&  0.71\\ 
\hline
\multicolumn{5}{c}{Model F}\\
\hline
F1  & 10.05 & 7.86 &0.041 &  0.82\\
\hline
\multicolumn{5}{c}{Model G}\\
\hline
G1  & 10.35 & 6.6 & 0.041 &  0.76\\
\hline\hline
\label{coreparams}
\end{tabular}
\end{table}

\subsection{Comparison to Previous Simulations}

Finally, four of the models (A, C, F, and  G) represent a set from which we can study the effect of changing various initial parameters. These simulations can also be directly compared to those of \citet{Basu2009b}. First, Models A and G represent extensions of the \citet{Basu2009b} models that each differ in a single property: Model A adds the effect of a step-like ionization profile while Model G adds the effect of ongoing low amplitude column density perturbations.

In Model A, the clump structures presented in Figure~\ref{fig:Av} are very similar to those presented in Figure 3 of \citet{Basu2009b}. Both show the onion like structure in the column density enhancement maps. However our clump regions are much larger since we start at a lower column density, with a consequent larger preferred fragmentation scale. The step-like ionization profile does however affect the simulation time, increasing it by almost a factor of 7/4 from $t_{\rm run}/t_{0} = 88$ to $t_{\rm run}/t_{0} = 143.6$. This is because at early times the low column density gas is almost flux-frozen, thus inhibiting the flow of neutral particles past the magnetic field. In Model G, the structures formed are almost identical to those depicted in \citet[][Figure 4]{Basu2009b}. The main difference between the two simulations is again the total run time. In this case, our model runs for $t_{\rm run}/t_{0} = 40$ compared to the $t_{\rm run}/t_{0} = 88$ for the equivalent \citet{Basu2009b} model. From this we can conclude that the addition of a step-like ionization profile serves to increase the collapse time of a cloud while the addition of ongoing perturbations has the opposite effect.

Models C and F differ from each other only in the initial value for the mass-to-flux ratio. Figure~\ref{fig:Av} (top right and bottom left) shows that this change has a significant effect on the evolution of the cloud. Due to the greater mass-to-flux ratio in Model F, the fragmentation scale is much smaller as predicted by linear analysis \citep{CB2006}. In addition, as discussed earlier, the corresponding time scale is also much shorter, resulting in a significantly shorter evolution time. Model F can also be compared to simulations presented by \citet{Basu2009b}.  Overall, this model produces structures similar to those found in the corresponding simulation in \citet{Basu2009b}, however the run time has been reduced from $t_{\rm run}/t_{0} = 23$ to $t_{\rm run}/t_{0} = 17$ due to the ongoing perturbations. Based on this difference in timescales, we can conclude that for supercritical mass-to-flux ratios, the effect of ongoing perturbations outweighs the effect of the step-like ionization profile. This is due to the fact that for supercritical values of $\mu$, the medium is unsupported against collapse. Perturbations to the column density only serve to make some regions of the medium even more supercritical and thus prone to collapse faster.

\section{Simulations in Context of the Two-Stage Fragmentation Model}
\label{simdiss}

BB12 presented a scenario in which an initially transcritical cloud can undergo an initial fragmentation event at low column densities and then a subsequent fragmentation event at higher column densities when the length and time scales for fragmentation undergo a significant decrease. This is the basis of the two-stage fragmentation model. The aim of the simulations presented in this paper is to determine the circumstances in which a cloud will undergo this two-stage fragmentation.
 
Of the seven models presented, four exhibit the features of the two-stage fragmentation scenario: Models B, C, D, and E. These models all assume a step-like ionization profile and transcritical mass-to-flux ratio. For all models, at early times, the initial column density within the region is very low ($A_{V,0} = 1 $ mag). As the cloud evolves, there will be a specific point in time where the column density and mass-to-flux ratio within a region correspond to the preffered length scale for collapse. This dependence on environmental parameters is clearly exhibited by the fragmentation times shown in Table~\ref{times}. For Model A, as shown by Figures~\ref{fig:max_sigma}~\&~\ref{fig:max_mtf}, the value of the column density enhancement and mass-to-flux ratio remain close to their initial values. It takes between 15 and 20 Myr for fragmentation to occur. When fragmentation does occur, the length scale is on the order of 10 pc, which corresponds to the predicted length scale from the linear model (see Figure 5 of BB12). In the other models, the time scale for fragmentation is much shorter. This is due to various reasons specific to each model. In Models B, C, D, and E, the ongoing perturbations cause both the column density and mass-to-flux ratio within some regions to increase. These increases work constructively to decrease the growth time. As such, the clouds in these models take less time to fragment initially as compared to Model A. Continuing on in the evolution, the other fragmentation event occurs after the ionization fraction has dropped in a step-like manner. The minimum growth time at this point is on the order of 1-2 Myr. In Table~\ref{times}, this time corresponds to the time between $t_{\rm clump}$ and $t_{\rm core}$. Looking at Models B, C, D, and E, we see that all exhibit roughly the predicted time frame for the growth of these core regions (see again Figure 5 of BB12). Compared to the other models, Models F and G follow much different paths. Model F only experienced one fragmentation event early on in the evolution. This is a direct consequence of the significantly supercritical initial mass-to-flux ratio. For Model G, we see that it also undergoes only single-stage fragmentation, and early on in the evolution. In this case, this is a direct consequence of the cosmic ray only ionization profile. 

In addition to the models discussed above, we ran two extra simulations to determine the effect of perturbing both the column density and magnetic field (i.e., maintaining a constant mass-to-flux ratio within each cell). These runs perturbed the two quantities every $t = 5 t_{0}$ and $t = 10 t_{0}$ respectively, similar to Models B and C. The results showed that by maintaining the mass-to-flux ratio within the cells, the evolution was slowed significantly and resembled that of Model A more than comparison models with only column density perturbations (i.e., Models B and C). These results agree with the enhanced ambipolar diffusion models of \citet{Zweibel2002} and \citet{FA2002} in that perturbations to the mass-to-flux ratio are required to speed up the evolution of the cloud.

\section{Summary}

We have performed thin disk non-ideal MHD simulations of molecular cloud collapse to test the parameter space required to form clusters via the two-stage fragmentation model presented in \citet{BB2012}. Some notable results are described below.

\begin{itemize}

\item The occurrence of two fragmentation events within the evolution of a cloud is highly dependent on the environment within which the region is evolving. Based on the simulations, we can say that the cloud must meet three criteria in order for it to experience two fragmentation events. First, the cloud must start out diffuse with an initially transcritical mass-to-flux ratio ($\mu_{0} \sim 1.1$). Second, the medium must have a step-like ionization profile. Finally, there must be some form of perturbations occurring within the region in order for structures formed from the second fragmentation event to remain distinct from each other.  

\item The clump and core regions formed in the simulations exhibit sizes and masses on the order of observations, with $\sim$~pc scale clumps of mass range 5~$M_{\odot}$~-~412~$M_{\odot}$ and $\sim$~0.1 pc scale cores of mass range 0.71~$M_{\odot}$~-~3.58~$M_{\odot}$.

\item The fragmentation is highly dependent on the initial environment of the cloud. Clouds with high initial ionization fractions (due to UV starlight) and transcritical mass-to-flux ratios will fragment into a few large structures while low ionization fraction (due to CR only) or super-critical initial mass-to-flux ratio clouds will fragment into many small structures.   

\item The mass-to-flux ratio maps show that the structure of the mass-to-flux ratio within the cloud is highly dependent on the environment. Regions with no perturbations (Model A) or a CR only ionization profile (Model G) yield maps where the mass-to-flux ratio traces the column density structure. Conversely, regions that undergo multiple perturbations and have a step-like column density dependent ionization fraction show very irregular mass-to-flux ratio maps. This is a consequence of both the perturbations and neutral-ion collision time. High ionization and frequent perturbations do not allow the mass to redistribute around the field lines. However, if given sufficient time, the mass-to-flux ratio will relax to a distribution that mimics the column density structure. 

\item The time scale for evolution of the cloud is highly dependent on the environment. Quiescent clouds evolve on much longer times scales than clouds experiencing ongoing perturbations. Our simulations also show that significant reductions to the evolutionary time requires perturbations to the mass-to-flux ratio via either the column density or magnetic field strength. Simulations that maintained the mass-to-flux ratio through the perturbations exhibit evolutionary time scales on the order of those observed in quiescent clouds. 

\item The evolutionary sequence in our models is consistent with long molecular cloud lifetimes ($\sim$ 30 Myr) in which most of the time is occupied by clump formation. The subsequent core and star formation phase occurs rapidly, on $\sim$ 2 Myr timescales, with core collapse occurring on even shorter timescales.
 
\end{itemize}

\section*{Acknowledgments}
We thank Chang-Goo Kim for valuable discussions. NDB was supported by a scholarship from the Natural Science and Engineering Research Council (NSERC) of Canada. SB was supported by a Discovery Grant from NSERC.


\begin{thebibliography}{38}
\expandafter\ifx\csname natexlab\endcsname\relax\def\natexlab#1{#1}\fi

\bibitem[{{Adams} \& {Shu}(2007)}]{AS2007}
{Adams}, F.~C. \& {Shu}, F.~H. 2007, ApJ, 671, 497

\bibitem[{{Bailey} \& {Basu}(2012)}]{BB2012}
{Bailey}, N.~D. \& {Basu}, S. 2012, ApJ, 761, 67

\bibitem[{{Basu} \& {Ciolek}(2004)}]{BC2004}
{Basu}, S. \& {Ciolek}, G.~E. 2004, ApJL, 607, L39

\bibitem[{{Basu} {et~al.}(2009{\natexlab{a}}){Basu}, {Ciolek}, {Dapp}, \&
  {Wurster}}]{Basu2009a}
{Basu}, S., {Ciolek}, G.~E., {Dapp}, W.~B., \& {Wurster}, J.
  2009{\natexlab{a}}, NewA, 14, 483

\bibitem[{{Basu} {et~al.}(2009{\natexlab{b}}){Basu}, {Ciolek}, \&
  {Wurster}}]{Basu2009b}
{Basu}, S., {Ciolek}, G.~E., \& {Wurster}, J. 2009{\natexlab{b}}, NewA, 14, 221

\bibitem[{{Benson} \& {Myers}(1989)}]{BM1989}
{Benson}, P.~J. \& {Myers}, P.~C. 1989, ApJS, 71, 89

\bibitem[{{Blitz} \& {Shu}(1980)}]{BS1980}
{Blitz}, L. \& {Shu}, F.~H. 1980, ApJ, 238, 148

\bibitem[{{Cho} \& {Kim}(2011)}]{Cho2011}
{Cho}, W. \& {Kim}, J. 2011, MNRAS, 410, L8

\bibitem[{{Ciolek} \& {Basu}(2006)}]{CB2006}
{Ciolek}, G.~E. \& {Basu}, S. 2006, ApJ, 652, 442

\bibitem[{{Fatuzzo} \& {Adams}(2002)}]{FA2002}
{Fatuzzo}, M. \& {Adams}, F.~C. 2002, ApJ, 570, 210

\bibitem[{{Frau} {et~al.}(2010){Frau}, {Girart}, {Beltr{\'a}n}, {Morata},
  {Masqu{\'e}}, {Busquet}, {Alves}, {S{\'a}nchez-Monge}, {Estalella}, \&
  {Franco}}]{Frau2010}
{Frau}, P., {Girart}, J.~M., {Beltr{\'a}n}, M.~T., {et~al.} 
2010, ApJ, 723, 1665

\bibitem[{{Froebrich} \& {Rowles}(2010)}]{Froebrich2010}
{Froebrich}, D. \& {Rowles}, J. 2010, MNRAS, 406, 1350

\bibitem[{{Gaustad}(1963)}]{Gaustad1963}
{Gaustad}, J.~E., 1963, ApJ, 138, 1050


\bibitem[{{Goldsmith} {et~al.}(2008){Goldsmith}, {Heyer}, {Narayanan}, {Snell},
  {Li}, \& {Brunt}}]{gol08}
{Goldsmith}, P.~F., {Heyer}, M., {Narayanan}, G., {et~al.} 
2008, ApJ, 680, 428

\bibitem[{{Hartmann}(2001)}]{Hartmann2001}
{Hartmann}, L. 2001, AJ, 121, 1030

\bibitem[{{Hartmann} {et~al.}(2001){Hartmann}, {Ballesteros-Paredes}, \&
  {Bergin}}]{HBB2001}
{Hartmann}, L., {Ballesteros-Paredes}, J., \& {Bergin}, E.~A. 2001, ApJ, 562,
  852

\bibitem[{{Hayashi}(1966)}]{Hayashi1966}
{Hayashi}, C. 1966, ARA\&A, 4, 171

\bibitem[{{Heiles} \& {Troland}(2005)}]{hei05}
{Heiles}, C. \& {Troland}, T.~H. 2005, ApJ, 624, 773

\bibitem[{{Hoyle}(1953)}]{Hoyle1953}
{Hoyle}, F. 1953, ApJ, 118, 513

\bibitem[{{Johnstone} {et~al.}(2004){Johnstone}, {Di Francesco}, \&
  {Kirk}}]{Johnstone2004}
{Johnstone}, D., {Di Francesco}, J., \& {Kirk}, H. 2004, ApJL, 611, L45

\bibitem[{{J{\o}rgensen} {et~al.}(2007){J{\o}rgensen}, {Johnstone}, {Kirk}, \&
  {Myers}}]{Jorgensen2007}
{J{\o}rgensen}, J.~K., {Johnstone}, D., {Kirk}, H., \& {Myers}, P.~C. 2007,
  ApJ, 656, 293

\bibitem[{{Kainulainen} {et~al.}(2009){Kainulainen}, {Beuther}, {Henning}, \&
  {Plume}}]{Kain2009}
{Kainulainen}, J., {Beuther}, H., {Henning}, T., \& {Plume}, R. 2009, A\&A,
  508, L35

\bibitem[{{Kirk} {et~al.}(2006){Kirk}, {Johnstone}, \& {Di
  Francesco}}]{Kirk2006}
{Kirk}, H., {Johnstone}, D., \& {Di Francesco}, J. 2006, ApJ, 646, 1009

\bibitem[{{Kudoh} \& {Basu}(2003)}]{kud03}
{Kudoh}, T. \& {Basu}, S. 2003, ApJ, 595, 842

\bibitem[{{Kudoh} \& {Basu}(2006)}]{kud06}
---. 2006, ApJ, 642, 270

\bibitem[{{Lee} \& {Myers}(2011)}]{LM2011}
{Lee}, C.~W. \& {Myers}, P.~C. 2011, ApJ, 734, 60

\bibitem[{{Larson}(1969)}]{Larson1969}
{Larson}, R.~B. 1969, MNRAS, 145, 271

\bibitem[{{McDaniel} \& {Mason}(1973)}]{MM1973}
{McDaniel}, E.~W. \& {Mason}, E.~A. 1973, The Mobility and Diffusion of Ions in
  Gases (New York: Wiley)

\bibitem[{{Mouschovias}(1977)}]{Mouschovias1977}
{Mouschovias}, T.~Ch. 1977, \apj, 211, 147

\bibitem[{{Murray}(2011)}]{Murray2011}
{Murray}, N. 2011, ApJ, 729, 133

\bibitem[{{Onishi} {et~al.}(1998){Onishi}, {Mizuno}, {Kawamura}, {Ogawa}, \&
  {Fukui}}]{oni98}
{Onishi}, T., {Mizuno}, A., {Kawamura}, A., {Ogawa}, H., \& {Fukui}, Y. 1998,
  ApJ, 502, 296

\bibitem[{{Onishi} {et~al.}(2002){Onishi}, {Mizuno}, {Kawamura}, {Tachihara},
  \& {Fukui}}]{Onishi2002}
{Onishi}, T., {Mizuno}, A., {Kawamura}, A., {Tachihara}, K., \& {Fukui}, Y.
  2002, ApJ, 575, 950

\bibitem[{{Pineda} {et~al.}(2010){Pineda}, {Goldsmith}, {Chapman}, {Snell},
  {Li}, {Cambr{\'e}sy}, \& {Brunt}}]{Pineda2010}
{Pineda}, J.~L., {Goldsmith}, P.~F., {Chapman}, N., {et~al.} 
2010, ApJ, 721, 686

\bibitem[{{Pinto} {et~al.}(2012){Pinto}, {Verdini}, {Galli}, \&
  {Velli}}]{Pinto2012}
{Pinto}, C., {Verdini}, A., {Galli}, D., \& {Velli}, M. 2012, A\&A, 544, A66

\bibitem[{{Rom{\'a}n-Z{\'u}{\~n}iga} {et~al.}(2009){Rom{\'a}n-Z{\'u}{\~n}iga},
  {Lada}, \& {Alves}}]{RZ2009}
{Rom{\'a}n-Z{\'u}{\~n}iga}, C.~G., {Lada}, C.~J., \& {Alves}, J.~F. 2009, ApJ,
  704, 183

\bibitem[{{Sadavoy} {et~al.}(2012){Sadavoy}, {di Francesco}, {Andr{\'e}},
  {Pezzuto}, {Bernard}, {Bontemps}, {Bressert}, {Chitsazzadeh}, {Fallscheer},
  {Hennemann}, {Hill}, {Martin}, {Motte}, {Nguyn Lu'O'Ng}, {Peretto}, {Reid},
  {Schneider}, {Testi}, {White}, \& {Wilson}}]{Sadavoy2012}
{Sadavoy}, S.~I., {di Francesco}, J., {Andr{\'e}}, P., {et~al.} 
2012, A\&A, 540, A10

\bibitem[{{Schiesser}(1991)}]{Schiesser1991}
{Schiesser}, W.~E. 1991, {The Numerical Method of Lines: Method of Integration
  of Partial Differential Equations} (Academic Press, San Diego)

\bibitem[{{Schmalzl} {et~al.}(2010){Schmalzl}, {Kainulainen}, {Quanz}, {Alves},
  {Goodman}, {Henning}, {Launhardt}, {Pineda}, \&
  {Rom{\'a}n-Z{\'u}{\~n}iga}}]{Schmalzl2010}
{Schmalzl}, M., {Kainulainen}, J., {Quanz}, S.~P., {et~al.} 
2010, ApJ, 725, 1327

\bibitem[{{Shampine}(1994)}]{Shampine1994}
{Shampine}, L.~F. 1994, {Numerical Solution of Ordinary Differential Equations}
  (Chapman \& Hall, New York)


\bibitem[{{Solomon} {et~al.}(1987){Solomon}, {Rivolo}, {Barrett} \&
  {Yahil}}]{Solomon1987}
{Solomon}, P.~M., {Rivolo}, A.~R., {Barrett}, J., \& {Yahil}, A. 1987, ApJ, 319, 741

\bibitem[{{Williams} {et~al.}(1994){Williams}, {de Geus}, \&
  {Blitz}}]{Williams1994}
{Williams}, J.~P., {de Geus}, E.~J., \& {Blitz}, L. 1994, ApJ, 428, 693

\bibitem[{{Xiang} {et~al.}(1984){Xiang}, {Liszt}, \& {Burton}}]{XLB1984}
{Xiang}, D.-L., {Liszt}, H.~S., \& {Burton}, W.~B. 1984, ChA\&A, 8, 105

\bibitem[{{Zucconi} {et~al.}(2001){Zucconi}, {Walmsley}, \& {Galli}}]{Zucconi2001}
{Zucconi}, A., {Walmsley}, C.~M., \& {Galli}, D. 2001, A\&A, 376, 650

\bibitem[{{Zweibel}(2002)}]{Zweibel2002}
{Zweibel}, E.~G. 2002, ApJ, 567, 962

\end{thebibliography}
\end{document}